\documentclass[aps,prd,twocolumn,10pt,superscriptaddress,nofootinbib,nobibnotes,longbibliography]{revtex4-1}

\usepackage{amssymb}
\usepackage{graphicx}
\usepackage{amsmath}
\usepackage{hyperref}
\usepackage{subfigure}
\usepackage{multirow}
\usepackage{setspace}
\usepackage{verbatim}
\usepackage{float}
\usepackage{color}
\usepackage{ulem}
\usepackage[utf8]{inputenc}
\usepackage[table,xcdraw]{xcolor}
\usepackage{makecell}

\begin{document}

\title{General bubble expansion at strong coupling}

\author{Jun-Chen Wang}
\email{junchenwang@stu.pku.edu.cn}
\affiliation{School of Physics, Peking University, Beijing 100871, China}

\author{Zi-Yan Yuwen}
\email{yuwenziyan@itp.ac.cn (Corresponding author)}
\affiliation{CAS Key Laboratory of Theoretical Physics, Institute of Theoretical Physics, Chinese Academy of Sciences (CAS), Beijing 100190, China}
\affiliation{School of Physical Sciences, University of Chinese Academy of Sciences (UCAS), Beijing 100049, China}

\author{Yu-Shi Hao}
\email{haoyushi@itp.ac.cn}
\affiliation{CAS Key Laboratory of Theoretical Physics, Institute of Theoretical Physics, Chinese Academy of Sciences (CAS), Beijing 100190, China}
\affiliation{School of Physical Sciences, University of Chinese Academy of Sciences (UCAS), Beijing 100049, China}

\author{Shao-Jiang Wang}
\email{schwang@itp.ac.cn (Corresponding author)}
\affiliation{CAS Key Laboratory of Theoretical Physics, Institute of Theoretical Physics, Chinese Academy of Sciences (CAS), Beijing 100190, China}
\affiliation{Asia Pacific Center for Theoretical Physics (APCTP), Pohang 37673, Republic of Korea}


\begin{abstract}
The strongly coupled system like the quark-hadron transition (if it is of first order) is becoming an active play yard for the physics of cosmological first-order phase transitions. However, the traditional field theoretic approach to strongly coupled first-order phase transitions is of great challenge, driving recent efforts from holographic dual theories with explicit numerical simulations. These holographic numerical simulations have revealed an intriguing linear correlation between the phase pressure difference (pressure difference away from the wall) to the nonrelativistic terminal velocity of an expanding planar wall, which has been reproduced analytically alongside both cylindrical and spherical walls from perfect-fluid hydrodynamics in our previous study but only for a bag equation of state. We also found, in our previous study, a universal quadratic correlation between the wall pressure difference (pressure difference near the bubble wall) to the nonrelativistic terminal wall velocity regardless of wall geometries. In this paper, we will generalize these analytic relations between the phase/wall pressure difference and terminal wall velocity into a more realistic equation of state beyond the simple bag model, providing the most general predictions so far for future tests from holographic numerical simulations of strongly coupled first-order phase transitions
\end{abstract}
\maketitle

\section{Introduction}\label{sec:introduction}

The cosmological first-order phase transition (FOPT)~\cite{Mazumdar:2018dfl,Hindmarsh:2020hop} is a quantum field analog of quantum tunneling in quantum mechanics and thermal transition in statistical mechanics. For a quantum field theory that exhibits a continuous symmetry breaking with the appearance of a potential barrier~\cite{Quiros:1999jp}, the cosmological FOPT occurs by randomly nucleating true-vacuum bubbles in the false-vacuum environment~\cite{Athron:2023xlk}, and then proceeds by accelerating expansion of bubble walls~\cite{Cai:2020djd,Lewicki:2022pdb} driven by the potential difference that is eventually balanced by the backreaction force during the asymptotic expansion stage~\cite{Wang:2022txy,Wang:2023kux}, and finally ends by violent bubble wall collisions~\cite{Jinno:2016vai,Jinno:2017fby,Konstandin:2017sat,Zhong:2021hgo} with longstanding bulk fluid motions afterward~\cite{Hindmarsh:2013xza,Hindmarsh:2015qta,Hindmarsh:2017gnf,Hindmarsh:2016lnk,Hindmarsh:2019phv,Guo:2020grp,Cai:2023guc,Sharma:2023mao,RoperPol:2023dzg}. The associated stochastic gravitational wave backgrounds (SGWBs)~\cite{Caprini:2015zlo,Caprini:2019egz}  and curvature perturbations~\cite{Liu:2022lvz} or even the primordial black holes~\cite{Liu:2021svg,Hashino:2021qoq,He:2022amv,Lewicki:2023ioy,Gouttenoire:2023naa,Baldes:2023rqv} provide comprehensive probes into our early Universe~\cite{Cai:2017cbj,Bian:2021ini,Caldwell:2022qsj}. 

Although much attention on cosmological FOPTs has focused on the model buildings and parameter space searching at the electroweak scales (see, for example,~\cite{Cai:2022bcf} and references therein) for their apparent advantage of promising detection in space-borne GW detectors, the current observational data has already manifested the potential power in constraining the cosmological FOPT at corresponding energy scales of PT much higher or lower than the electroweak scales. For example, with the first three observing runs of Advanced LIGO-Virgo' data, the strong super-cooling FOPTs at LIGO-Virgo band have been marginally ruled out~\cite{Yu:2022xdw} when both contributions from wall collisions and sound waves are present as a general improvement to the previous works~\cite{Romero:2021kby,Huang:2021rrk,Jiang:2022mzt,Badger:2022nwo} with a single source. In particular, recent detection of SGWBs from the pulsar-timing-array (PTA) observations~\cite{Xu:2023wog,NANOGrav:2023gor,EPTA:2023sfo,Reardon:2023gzh} has renewed the interest in strongly coupled system like the quark-gluon/hadron PT at the quantum chromodynamics (QCD) scales.

The cosmological PT of a strongly coupled system, if it is of first-order, has thus became an alternative probe in addition to the traditional heavy-ion collisions and lattice simulations for investigating the strong dynamics in QCD physics from various cosmological observations like the recent PTA constraints~\cite{NANOGrav:2023hvm,Addazi:2023jvg,Athron:2023mer,Fujikura:2023lkn,Han:2023olf,Franciolini:2023wjm,Bian:2023dnv,Jiang:2023qbm,Ghosh:2023aum,Xiao:2023dbb,Li:2023bxy,DiBari:2023upq,Cruz:2023lnq,Wu:2023hsa,Du:2023qvj,Gouttenoire:2023bqy,Ahmadvand:2023lpp,Wang:2023bbc,He:2023ado} at QCD scales. In particular, the PTA constraint~\cite{Gouttenoire:2023bqy} on the FOPT at QCD scales allows for the productions of  solar-mass primordial black holes (PBHs)~\cite{Liu:2021svg}, which, however, might be disfavoured by the accompanying constraints from curvature perturbations~\cite{Liu:2022lvz} as shown specifically for a holographic QCD model~\cite{He:2023ado}. On the other hand, the strongly coupled FOPT can in return serve as a play yard for exploring the nonequilibrium physics of cosmological FOPT. However, unlike the usual weakly coupled FOPT, the strongly coupled FOPT is difficult to deal with from the traditional perturbative field-theory approach due to its strong-dynamics nature.

Nevertheless, the holographic principle, especially the AdS/CFT correspondence~\cite{Maldacena:1997re,Witten:1998qj,Gubser:1998bc} as a specific realization of the strong-weak duality, can be naturally applied to the strongly coupled FOPT in recent studies on bubble nucleation~\cite{Bigazzi:2020avc,Ares:2020lbt,Bigazzi:2020phm,Zhu:2021vkj,Ares:2021ntv,Ares:2021nap,Morgante:2022zvc} and bubble expansion~\cite{Bea:2021zsu,Bigazzi:2021ucw,Janik:2022wsx,Bea:2022mfb} as well as bubble-collision phenomenology~\cite{Cai:2022omk,He:2022amv,Zhao:2022uxc,Li:2023mpv,He:2023ado}. In particular, the numerical simulations~\cite{Bea:2021zsu,Janik:2022wsx} from two very different holographic models reveal a similar linear correlation between the phase pressure difference\footnote{Note that, owing to the presence of sound shell with a nonvanishing fluid-velocity profile around the bubble wall, the phase pressure difference is different from the wall pressure difference, the former of which takes the pressure difference away from the bubble wall (in fact, away from the sound shell) while the latter of which takes the pressure difference near the bubble wall.} and the terminal velocity of an expanding planar wall as also derived analytically from a nonperturbative top-down approach~\cite{Bigazzi:2021ucw}. However, such a correlation has not been explored yet in the holographic numerical simulation~\cite{Bea:2022mfb} for a cylindrical wall due to the high costs of computational power. Based on the same reason, the holographic numerical simulation has also not been conducted to date for the more realistic case of spherical wall expansion.

Intriguingly, besides the linear correlation between the phase pressure difference and terminal planar-wall velocity, the holographic numerical simulations~\cite{Bea:2021zsu,Janik:2022wsx} have also unfolded two characteristic features for the strongly coupled FOPT: (i) The terminal wall velocity is marginally nonrelativistic. This can be understood that, as the bubble wall strongly interacts with the ambient plasma, the backreaction force is so rapidly growing that it only takes a very short time duration for the accelerating expansion stage until the backreaction force could balance the driving force. Hence, the strong dynamics can force the bubble wall to quickly saturate at a small velocity; (ii) The perfect-fluid hydrodynamics works extremely well in the whole range of bubble expansion except at the wall position. This can be understood as the bubble wall now moves so slowly (nonrelativistically) that the particles have enough time to fully thermalize before the bubble wall has swept over. Hence, the strong dynamics can also help to establish perfect-fluid hydrodynamic approximation except at the wall. Note that with appropriate junction conditions across the bubble wall, the perfect-fluid hydrodynamic approximation might as well work effectively at the wall position~\cite{Wang:2022txy,Wang:2023kux}.

The above-mentioned nonrelativistic terminal wall velocity and perfect-fluid hydrodynamics approximation revealed by the holographic numerical simulations for the strongly coupled FOPT have indicated that it might be feasible to derive the linear correlation between the phase pressure difference and terminal planar-wall velocity from bottom-up approach by fully appreciating the perfect-fluid hydrodynamics in the nonrelativistic limit of a planar-wall expansion. This is what we achieved in Ref.~\cite{Li:2023xto} not only for the planar wall but also for both cylindrical and spherical walls provided with a bag equation of state (EOS).

However, in both holographic numerical simulations and realistic models of strongly coupled FOPTs, the EOS cannot be fixed exactly by the bag model. It is therefore necessarily useful to generalize our previous study~\cite{Li:2023xto} directly into the case beyond the bag EOS, and in particular, to provide analytic approximations for practical use without going over again the whole numerical evaluations. We therefore first set up the conventions and requisite formulas for later use in Sec.~\ref{sec:setup}, and then derive in the nonrelativistic wall limit for its correlations to the phase pressure difference and wall pressure difference in Sec.~\ref{sec:phasedeltap} and Sec.~\ref{sec:walldeltap}, respectively. Finally, the Sec.~\ref{sec:condis} is devoted to conclusions and discussions. Appendix~\ref{app:numodel} is provided for a self-containing introduction to the hydrodynamics beyond the bag EOS.

\section{Strongly coupled FOPT}\label{sec:setup}

In this section, we will introduce the necessary notations and conventions closely following Ref.~\cite{Wang:2023kux} in order to generalize the results of our previous study~\cite{Li:2023xto}.

For a generally coupled system of scalar field and thermal plasma, the joined dynamics is governed by a series of Boltzmann equations for the distribution functions of each species. By considering the late stage of a fast FOPT, one can take advantage of simplifications from the flat-spacetime background, self-similar expansion, thin-wall geometry, and steady-state evolution. Therefore, the scalar-plasma system can be further reduced into a wall-fluid system~\cite{Wang:2023kux} that can be well described by the perfect-fluid hydrodynamics with corresponding energy-momentum tensor of form
\begin{align}
T^{\mu\nu}=(e+p)u^\mu u^\nu+p\eta^{\mu\nu},
\end{align}
where $e$, $p$ are the total energy density and pressure, and $u^\mu\equiv\mathrm{d}x^\mu/\mathrm{d}\tau$ is the four velocity of the fluid element at $x^\mu\equiv(t,z,x=0,y=0), (t,\rho,\varphi=0,z=0), (t,r,\theta=0,\varphi=0)$ for planar, cylindrical, and spherical wall geometries, respectively. Here, the corresponding coordinate systems are established at the center of the bubble in such a way that the fluid element only moves in the $x^1$ direction with the other two spatial directions fixed constantly, for example, all at zero. Hence, the four velocity of bulk fluids also reads $u^\mu=\gamma(v)(1,v,0,0)$ from the three velocity $v\equiv\mathrm{d}x^1/\mathrm{d}x^0$ via the Lorentz factor $\gamma(v)\equiv1/\sqrt{1-v^2}$. The similarity of bubble expansion during the asymptotic stage at late time preferentially defines a convenient self-similar coordinate system $(T=t, X=x^1/x^0\equiv\xi)$ so that $v(\xi)$ traces the fluid velocity at $x^1=\xi t$ in the background plasma frame. Besides, the steady-state expansion of the thin wall also preferentially defines an observer frame comoving with the wall at $x_w^1(x^0)=\xi_w t$ traced by the wall velocity $\xi_w$. Hence, in the comoving wall frame, the bulk-fluid four velocity reads $u^\mu=\bar{\gamma}(1,-\bar{v},0,0)$ with $\bar{\gamma}\equiv\gamma(\bar{v})=1/\sqrt{1-\bar{v}^2}$, where the negative sign before the wall-frame three velocity $\bar{v}=(\xi_w-v)/(1-\xi_w v)\equiv\mu(\xi_w,v)$ is introduced to ensure a positive $\bar{v}$ for later convenience. Here, the abbreviation $\mu(\zeta, v(\xi))\equiv(\zeta-v)/(1-\zeta v)$ denotes for the Lorentz boost of the bulk fluid velocity $v(\xi)$ in the background plasma frame into a $\zeta$-frame velocity seen in the comoving frame with velocity $\zeta$.

With the wall-fluid approximation for the coupled scalar-plasma system of cosmological FOPTs, the equation of motions (EoMs) of the wall-fluid system is given by the conservation of the total energy-momentum tensor $\nabla_{\mu}T^{\mu\nu}=0$, which can be projected parallel along and perpendicular to the bulk fluid direction~\cite{Espinosa:2010hh} that can be further combined into following equations for the profiles of fluid velocity $v(\xi)$ and total enthalpy $w(\xi)=e+p$,
\begin{align}
 D\frac{v}{\xi}&=\gamma(v)^2(1-\xi v)\left(\frac{\mu(\xi,v)^2}{c_s^2}-1 \right)\frac{\text{d}v}{\text{d}\xi}, \label{eq:dv}\\
 \frac{\text{d}\ln w}{\text{d}\xi}&=\gamma(v)^2\mu(\xi,v)\left(\frac{1}{c_s^2}+1\right)\frac{\text{d}v}{\text{d}\xi}. \label{eq:dw}
\end{align}
Here $D=0,1,2$ correspond to planar, cylindrical, and spherical walls~\cite{Leitao:2010yw}, respectively, and the sound velocity $c_s = \sqrt{\partial_{\xi}p/\partial_{\xi}e}$ should be in general a function of $\xi$~\cite{Wang:2022lyd}. To further maintain the conservation of total energy-momentum tensor across the discontinuous interfaces at the bubble wall $\xi=\xi_w$ and shockwave front $\xi=\xi_{sh}$, appropriate junction conditions should be imposed from the temporal and spatial components of $\nabla_{\mu}T^{\mu\nu}=0$. Specifically, in the comoving frame of the bubble wall, the following junction conditions,
\begin{align}
 w_-\overline{\gamma}_-^2\overline{v}_-&=w_+\overline{\gamma}_+^2\overline{v}_+,\label{eq:WallJunction1}\\
 w_-\overline{\gamma}^2_-\overline{v}_-^2+p_-&=w_+\overline{\gamma}^2_+\overline{v}_+^2+p_+, \label{eq:WallJunction2}
\end{align}
are hold across the bubble wall, where $w_{\pm}$, $p_{\pm}$, $\overline{v}_{\pm}$, and $\overline{\gamma}_{\pm}\equiv \gamma(\overline{v}_{\pm})$ are the enthalpy, pressure, wall-frame fluid velocity, and corresponding Lorentz factors just right in front and back of the bubble wall, respectively. Besides, in the comoving frame of the shockwave front, the following junction conditions,
\begin{align}
w_L\tilde{\gamma}_L^2\tilde{v}_L&=w_R\tilde{\gamma}_R^2\tilde{v}_R,\label{eq:ShockJunction1}\\
w_L\tilde{\gamma}^2_L\tilde{v}_L^2+p_L&=w_R\tilde{\gamma}^2_R\tilde{v}_R^2+p_R,\label{eq:ShockJunction2}
\end{align}
are hold across the shockwave front, where $w_{R/L}$, $p_{R/L}$, $\tilde{v}_{R/L}$, and $\tilde{\gamma}_{R/L}\equiv \gamma(\tilde{v}_{R/L})$ are the enthalpy, pressure, shock-frame fluid velocity, and corresponding Lorentz factors just right in front and back of the shockwave front. Therefore, the combination of the fluid EoMs~\eqref{eq:dv} and~\eqref{eq:dw} with the junction conditions~\eqref{eq:WallJunction1},~\eqref{eq:WallJunction2},~\eqref{eq:ShockJunction1}, and~\eqref{eq:ShockJunction2} together ensures the conservation of total energy-momentum tensor in the whole range of the fluid profile.

The fluid EoMs~\eqref{eq:dv} and~\eqref{eq:dw} can be readily solved numerically for the detonation and deflagration modes with the junction condition~\eqref{eq:WallJunction1} at the bubble wall and junction condition~\eqref{eq:ShockJunction1} at the shockwave front (if any) provided with an extra assumption on the EOS. For a strongly coupled FOPT, the MIT bag EOS~\cite{Chodos:1974je} is usually assumed as a good approximation with corresponding sound velocity $c_s=1/\sqrt{3}$ independent of $\xi$. A more general EOS dubbed $\nu$-model~\cite{Leitao:2014pda} renders two constant sound velocities $c_\pm^2=\partial_\xi p_\pm/\partial_\xi e_\pm$ outside and inside of the bubble wall, respectively, where 
\begin{align}
e_{\pm}&=a_{\pm}T_{\pm}^{\nu_{\pm}}+V_0^{\pm},\label{eq:nue}\\
p_{\pm}&=c_{\pm}^2a_{\pm}T_{\pm}^{\nu_{\pm}}-V_0^{\pm}\label{eq:nup}
\end{align}
are the total energy density and pressure just right in the front and back of the bubble wall, respectively. Here, $V_0^\pm\equiv V_0(\phi_\pm)$ is the zero-temperature part of total effective potential $V_\mathrm{eff}(\phi,T)=V_0(\phi)+\Delta V_T(\phi,T)$ at the false and true vacua $\phi_\pm$, respectively. It is easy to see that $\nu_{\pm}= 1+1/c_{\pm}^2$. With above $\nu$-model EOS, the wall-frame fluid velocities $\bar{v}_\pm$ from the junction conditions~\eqref{eq:WallJunction1} and~\eqref{eq:WallJunction2} can be related by
\begin{equation}
    \overline{v}_{+}=\frac{1}{1+\alpha_{+}}\left( qX_{+}\pm \sqrt{q^2X_{-}^2+\alpha_{+}^2+(1-c_{+}^2)\alpha_+ +q^2c_{-}^2-c_{+}^2} \right),
    \label{eq:vbarpm}
\end{equation}
with abbreviations $q\equiv(1+c_{+}^2)/(1+c_{-}^2)$, $X_{\pm}\equiv\overline{v}_{-}/2\pm c_{-}^2/(2\overline{v}_{-})$, $\alpha_{+}\equiv \Delta V_0 /(a_+ T_+^{\nu_+})$, and $\Delta V_0 \equiv V_0^{+}-V_0^{-}$. One can also define the strength factor $\alpha_{N}\equiv \Delta V_0 /(a_+ T_N^{\nu_+})$ at null infinity $\xi=1$ (unperturbed by fluid motions) so that $\alpha_{+}w_{+}=\alpha_{N}w_{N}=(1+c_{+}^2)\Delta V_0$. The hydrodynamic solutions for the above $\nu$-model EOS can be solved numerically in Appendix~\ref{app:numodel}.

To see the nonrelativistic behavior of the phase pressure difference (driving force per unit area) between the innermost and outermost parts of the fluid profile~\cite{Wang:2022txy,Wang:2023kux},
\begin{align}
p_{\rm dr}&=\Delta V_{\rm eff}=-\Delta p=\Delta (-c_s^2aT^{\nu}+V_0)\nonumber\\
&=-\frac{c_{+}^2}{1+c_{+}^2}w_{N}+\frac{c_{-}^2}{1+c_{-}^2}w_{O}+\frac{1}{1+c_{+}^2}\alpha_{N}w_{N},
\end{align}
we consider the deflagration expansion of bulk fluid with a compressive shockwave as a sound shell in front of the bubble wall, in which case we can equal the enthalpy at null infinity $w_N\equiv w(\xi=1)=w(\xi=\xi_{sh}+0^+)\equiv w_R$ to the enthalpy just in front of the shockwave front, and equal the enthalpy at the origin $w_O\equiv w(\xi=0)=w(\xi=\xi_w+0^-)\equiv w_-$ to the enthalpy just behind the bubble wall. Further note that $w_-$ can be even reduced to depend only on $\xi_w$, $v_+$, and $w_{+}$ by adopting the junction condition~\eqref{eq:WallJunction1} with $\overline{v}_{+}=\mu(\xi_w,v_+)$ and $\overline{v}_{-}=\xi_w$, where $w_+$ can be further expressed in terms of $\xi_w$, $v_+$, and observable parameters at null infinity like $\alpha_N$ and $\omega_N$ by
\begin{equation}
    w_{+}=\frac{(1+c_{-}^2)(1-v_{+}^2)\xi_{w}\alpha_{N}w_{N}}{c_{+}^2(\xi_{w}+c_{-}^2v_{+})(1-v_{+}\xi_{w})-(c_{-}^2+v_{+}\xi_{w})(\xi_{w}-v_{+})}
    \label{eq:EnthalpyPlus}
\end{equation}
from the minus-sign branch of~\eqref{eq:vbarpm}. Now the phase pressure difference reads purely in terms of the sound velocities $c_\pm$, null-infinity observables $\alpha_N$ and $w_N$, bubble wall velocity $\xi_w$, and fluid velocity $v_+$ (to be determined later),
\begin{align}
  \label{eq:DpON}
  &\frac{p_{\rm dr}}{w_{N}}=\left[ \frac{1}{1+c_{+}^2}\right.\nonumber\\
  &\left.+\frac{c_{-}^2(\xi_{w}-v_{+})(1-v_{+}\xi_{w})}{c_{+}^2(\xi_{w}+c_{-}^2v_{+})(1-v_{+}\xi_{w})-(c_{-}^2+v_{+}\xi_{w})(\xi_{w}-v_{+})} \right]\alpha_{N}\nonumber\\
  &-\frac{c_{+}^2}{1+c_{+}^2},
\end{align}
where $v_{+}=\mu(\xi_w,\overline{v}_{+}(\xi_w,\alpha_+))$ from the minus-sign branch of~\eqref{eq:vbarpm} can be further reduced in terms of $\xi_w$ and $\alpha_+$. Therefore, as long as we can find a relation between $\alpha_{+}$ and $\alpha_N$, which can be achieved approximately to the leading order (LO) in $\xi_w$ for planar, cylindrical, and spherical walls, we can eventually arrive at the direct relation between the phase pressure difference $p_{\rm dr}$ and bubble wall velocity $\xi_w$ solely in terms of the $\nu$-model EOS $c_\pm$ and null-infinity observables $\alpha_N$ and $w_N$ without reference to the underlying microscopic physics.

\section{Phase pressure difference}\label{sec:phasedeltap}

In this section, we analytically derive the approximated relation between the phase pressure difference $p_{\rm dr}=\Delta V_{\rm eff}=p_O-p_N$ and the bubble wall velocity $\xi_w$ with $\nu$-model EOS for planar, cylindrical, and spherical wall geometries.

\subsection{Planar wall}

\begin{figure*}
\centering
\includegraphics[width=0.49\textwidth]{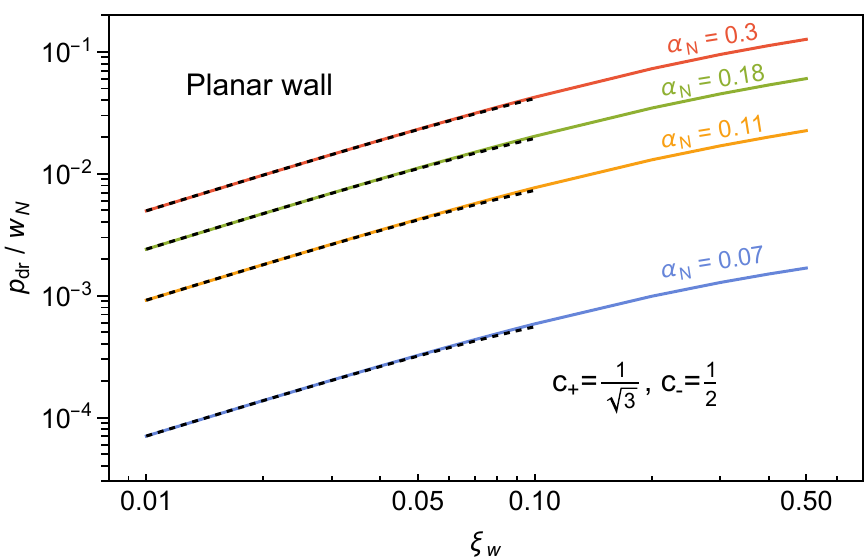}
\includegraphics[width=0.49\textwidth]{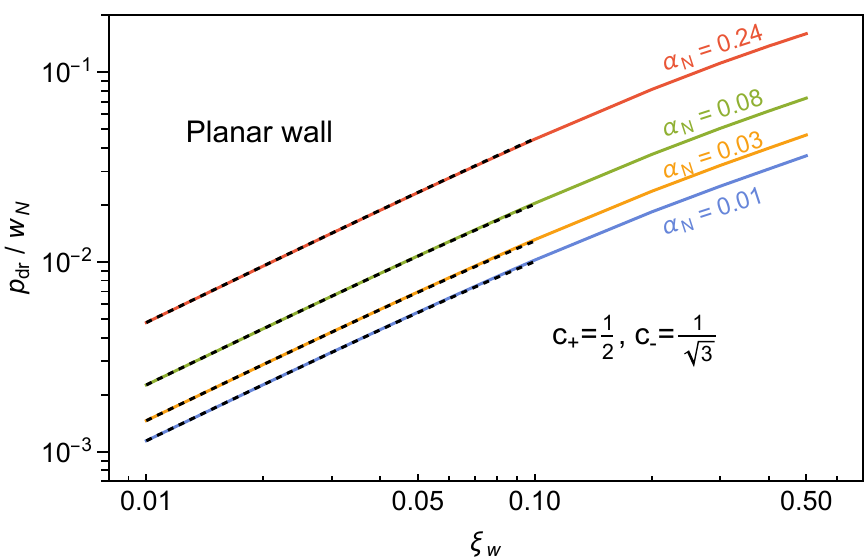}\\
\includegraphics[width=0.49\textwidth]{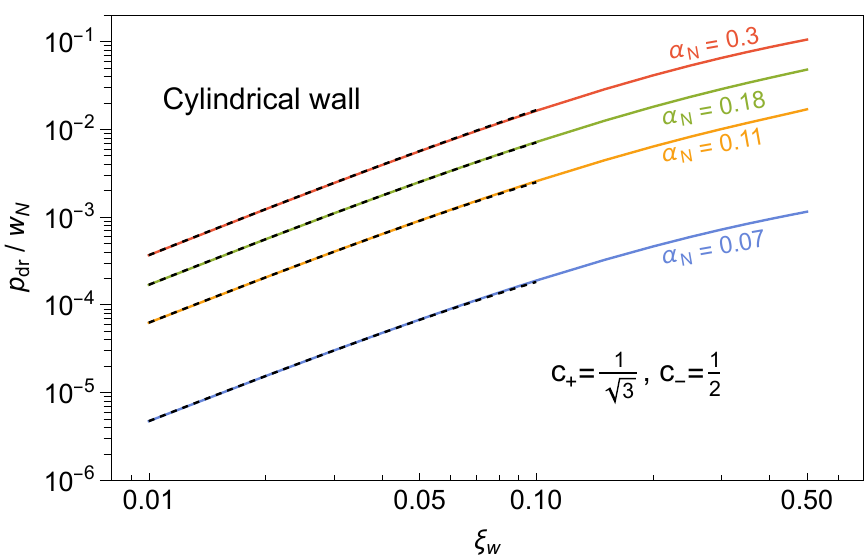}
\includegraphics[width=0.49\textwidth]{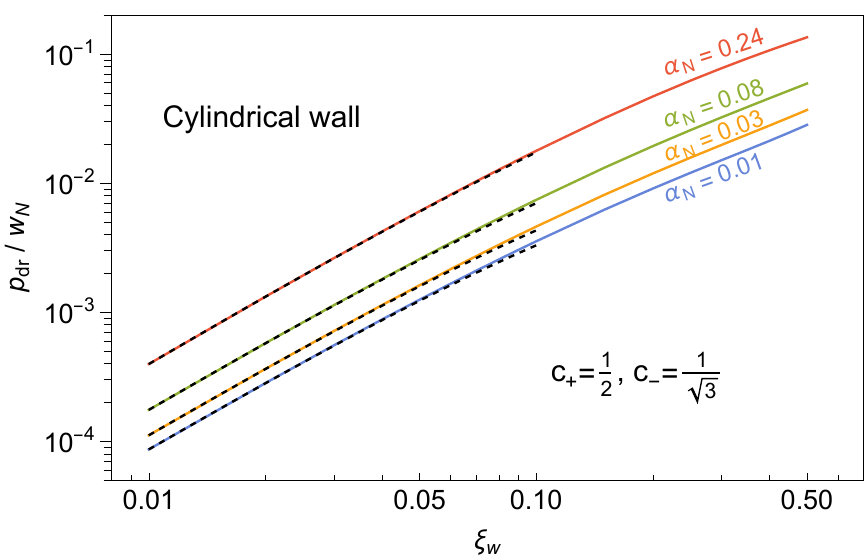}\\
\includegraphics[width=0.49\textwidth]{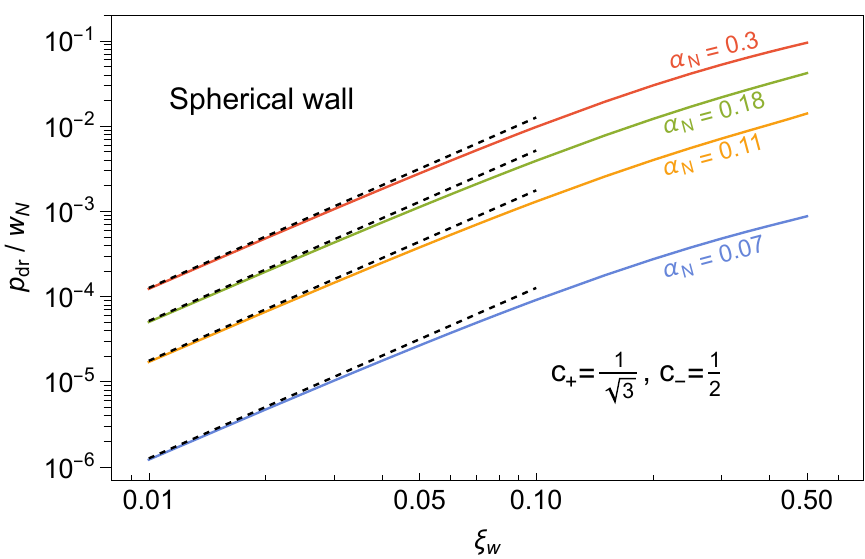}
\includegraphics[width=0.49\textwidth]{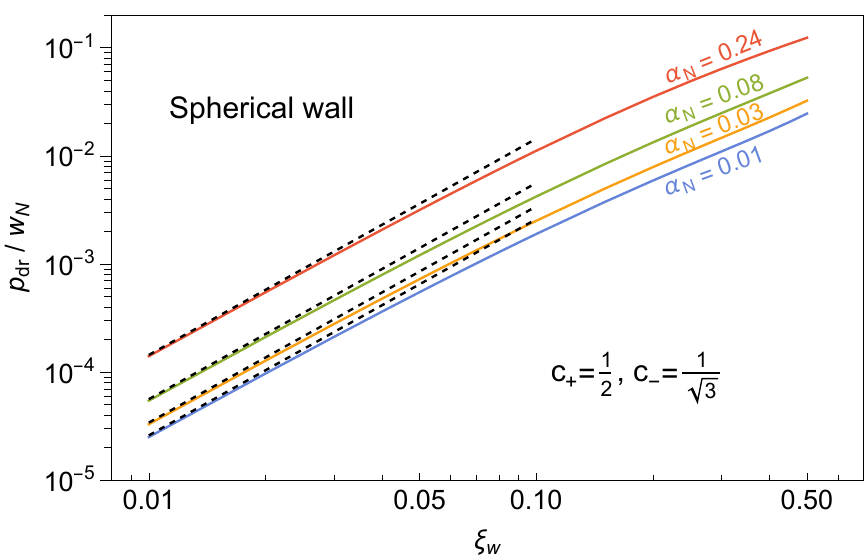}\\
\caption{The comparison between our analytical approximations (dashed lines) and the exact numerical evaluations (solid lines) for the relations~\eqref{planar:pdrLO} (top), \eqref{cylindrical:pdrLO} (middle), and~\eqref{spherical:pdrLO} (bottom) between the phase pressure difference $p_{\rm dr}/w_N$ and wall velocity $\xi_w$ given some illustrative values for the asymptotic strength factor $\alpha_N$ in both cases of $\nu$-model EOS with sound velocities $c_+>c_-$ (left) and $c_+<c_-$ (right).}
\label{fig:phase}
\end{figure*}

For a planar wall, the nonvanishing fluid profile is depicted by the fluid EoM~\eqref{eq:dv} with $D=1$,
\begin{equation}
    \label{planar:EOM}
    (\mu(\xi,v)^2-c_+^2)\frac{\text{d}v}{\text{d}\xi}=0,
\end{equation}
whose solutions are either $\text{d}v/\text{d}\xi=0$, namely, $v=\text{const}$, or $\mu(\xi,v)=c_+$, which would lead to $\xi>c_+$ for $v>0$ but with no deflagration regime. Hence, the only solution should be $v=\text{const.}=v_+$ in the sound shell and the corresponding enthalpy profile from~\eqref{eq:dw} with $\text{d}v/\text{d}\xi=0$ also stays constant in the sound shell, $w_+=\text{const.}=w_{L}$. This $w_{L}$ can be related to $w_R=w_N$ by the junction condition~\eqref{eq:ShockJunction1} via $\tilde{v}_{R}=\xi_{sh}$ and $\tilde{v}_L=\mu(\xi_{sh},v_{sh})$ from the fluid velocity $v_{sh}\equiv v(\xi_{sh}+0^-)$ just behind the shockwave front $\xi_{sh}$. To further determine $v_{sh}$ and $\xi_{sh}$, note that the constant velocity profile in the sound shell implies $v_{sh}=v_+=\mu(\xi_w,\overline{v}_+(\overline{v}_-,\alpha_+))$ with $\overline{v}_-=\xi_w$ and $\overline{v}_+(\overline{v}_-,\alpha_+)$ given by the
minus-sign branch of~\eqref{eq:vbarpm}. Thus, $v_{sh}$ can be expressed in terms of $\xi_w$ and $\alpha_+$ alone. Once $v_{sh}$ is determined, $\xi_{sh}$ can be directly obtained from the shock front condition $\mu(\xi_{sh},v_{sh})\xi_{sh}=c_+^2$. Hence, $\alpha_N/\alpha_+=w_+/w_N$ can be
derived in terms of $\xi_w$ and $\alpha_+$ alone, which can be expanded as
\begin{align}
    \frac{\alpha_N}{\alpha_+}=1&+\frac{c_-^2-c_+^2+(1+c_-^2)\alpha_+}{c_-^2c_+}\xi_{w}\nonumber\\
    &+\frac{[c_-^2-c_+^2+(1+c_-^2)\alpha_+]^2}{2c_-^4c_+^2}\xi_w^2+\mathcal{O}(\xi_w^3).
    \label{planar:alphaLO}
\end{align}
We can reverse the above relation to get $\alpha_+$ expressed in terms of $\xi_w$ and $\alpha_N$. Then, we can plug $\alpha_{+}(\xi_w,\alpha_N)$ into the minus-sign branch of~\eqref{eq:vbarpm} to get $\overline{v}_+(\overline{v}_-\equiv \xi_w,\alpha_{+}(\xi_w,\alpha_N))$. Next, we can further expand $v_+=\mu(\xi_w,\overline{v}_+(\xi_w,\alpha_N))$ in $\xi_w$, which finally yields the phase pressure difference~\eqref{eq:DpON} in the small $\xi_w$ limit up to the next-to-leading order (NLO) as
\begin{align}
    \frac{p_{\rm dr}}{w_N}=&\frac{c_+[c_{-}^2-c_+^2  + (1+c_-^2)\alpha_N]}{c_-^2(1+c_+^2)}\xi_{w}\nonumber\\
    &+\frac{c_{-}^2-c_+^2  + (1+c_-^2)\alpha_N}{2c_-^4(1+c_+^2)^2}\left[\alpha_N-c_+^4-c_+^2(3+\alpha_N)\right.\nonumber\\
    &\left.+c_-^2(1-c_+^2)(1+\alpha_N)\right]\xi_w^2+\mathcal{O}(\xi_w^3).
    \label{planar:pdrLO}
\end{align}
In the bag limit $c_\pm=c_s$, this analytic approximation reduces to the same linear correlation $p_\mathrm{dr}=\alpha_Nw_N\xi_w/c_s+\mathcal{O}(\xi_w^2)$ at the leading order as our previous estimation~\cite{Li:2023xto}. To see the goodness of fit for our analytical approximation, we can separately evaluate the phase pressure difference numerically from the exact numerical solutions, and then find a perfect match for both cases with $c_+>c_-$ and $c_+<c_-$ at NLO as shown in Fig.~\ref{fig:phase}. This leading-order linear dependence in the planar-wall velocity can be tested explicitly in Sec.~\ref{sec:condis} with respect to the holographic numerical simulation of a strongly coupled FOPT with a planar wall~\cite{Bea:2021zsu}.

\subsection{Cylindrical wall}

For a cylindrical wall with $D=1$, the fluid EoM~\eqref{eq:dv} to the order of $v^2$,
\begin{equation}
    \label{cylindrical:EOM}
    \frac{\text{d} v}{\text{d} \xi}=\frac{c_+^2 v}{\xi(\xi^2-c_+^2)}-\frac{(c_+^2+\xi^2-2)c_+^2v^2}{(c_+^2-\xi^2)^2}+\mathcal{O}(v^3),
\end{equation}
can be solved as
\begin{align}
v(\xi)&=\frac{c_+(c_+^2-\xi^2)}{\xi}\bigg/\left[2c_+(c_+^2-1)+\frac{1}{v_+\xi_w}\sqrt{\frac{c_+^2-\xi^2}{c_+^2-\xi_w^2}}\text{Sol}\right],\label{cylindrical:solution}\\
\text{Sol}&=c_+\xi_w(2v_+-\xi_w)+c_+^3(1-2v_+\xi_w)\nonumber\\
&+v_+\xi_w(c_+^2-2)\sqrt{c_+^2-\xi_w^2}\ln\left(\frac{\xi \left(c_++\sqrt{c_+^2-\xi_w^2}\right)}{\xi_w \left(c_++\sqrt{c_+^2-\xi^2}\right)}\right)\nonumber
\end{align}
given the condition $v(\xi_w+0^+)=v_+$ at the bubble wall. It is easy to see from~\eqref{cylindrical:solution} that the shock front where $v(\xi)$ drops to zero is now approximated at $\xi=\xi_{sh}=c_+$ with $w(c_+)=w_N$, from which we can integrate the fluid EoM~\eqref{eq:dw} to evaluate $w_+$ at $\xi_w$ from $\text{d} \ln w/ \text{d} \xi$ as estimated shortly below. To estimate $\text{d} \ln w/ \text{d} \xi$, we first insert~\eqref{cylindrical:solution} into~\eqref{eq:dw} and then expand $\text{d} \ln w/ \text{d} \xi$ to the order of $v_+^2$. Hence, $\alpha_N/ \alpha_+=w_+/w_N$ is now a function of $\xi_w$, $\alpha_+$, and $v_+=\mu(\xi_w,\overline{v}_+)$. After inserting $\overline{v}_{+}(\xi_w,\alpha_+)$ from the minus-sign branch of~\eqref{eq:vbarpm}, $\alpha_N/ \alpha_+$ can be expanded in the small $\xi_w$ limit as
\begin{align}
 \frac{\alpha_N}{\alpha_+}=1+&\frac{c_-^2-c_+^2+(1+c_-^2)\alpha_+}{2c_-^4c_+^2(1+c_+^2)}\left[c_+^2-c_-^2-(1+c_-^2)\alpha_+\right.\nonumber\\
 &\left.+2c_-^2(1+c_+^2)\ln \frac{2c_+}{\xi_w}\right]\xi_w^2+\mathcal{O}(\xi_w^4).
 \label{cylindrical:alphaLO}
\end{align}
Reversing the above relation to get $\alpha_{+}(\xi_w,\alpha_N)$ and plugging it into the minus-sign branch of~\eqref{eq:vbarpm}, we can derive $\overline{v}_{+}(\xi_w,\alpha_N)$ as a function of $\xi_w$ and $\alpha_N$. We next further expand $v_+=\mu(\xi_w,\overline{v}_{+}(\xi_w,\alpha_N))$ in terms of $\xi_w$ and then insert it into~\eqref{eq:DpON}, finally the phase pressure difference can be obtained in the small $\xi_w$ limit as 
\begin{align}
 &\frac{p_{\rm dr}}{w_N}=\frac{c_-^2-c_+^2+(1+c_-^2)\alpha_N}{2c_-^4(1+c_+^2)^2}\left[(1+c_-^2)\alpha_N\right.\nonumber\\
 &\left.-(c_+^2+2c_+^2c_-^2+c_-^2)+2c_-^2(1+c_+^2)\ln \frac{2c_+}{\xi_w}\right]\xi_w^2+\mathcal{O}(\xi_w^4).
    \label{cylindrical:pdrLO}
\end{align}
Note that the purely quadratic term in $\xi_w$ in the above approximation is actually an NLO term, while the term proportional to $\xi_w^2\ln\xi_w$ is at the leading order as it is larger than the purely quadratic term in $\xi_w$.
This analytic expression serves as an even better approximation in the bag limit $c_\pm\to c_s=1/\sqrt{3}$ compared to our previous estimation~\cite{Li:2023xto}, and also perfectly matches the exact numerical evaluation as shown in Fig.~\ref{fig:phase} for both cases with $c_+>c_-$ and $c_+<c_-$, where the distinctive logarithmic dependence can be directly tested in future holographic numerical simulations of strongly coupled FOPTs with a cylindrical wall~\cite{Bea:2022mfb}.

\subsection{Spherical wall}

For a spherical wall with $D=2$, the fluid EoM~\eqref{eq:dv} to the order of $v^2$,
\begin{equation}
    \label{spherical:EOM}
    \frac{\text{d} v}{\text{d} \xi}=\frac{2c_+^2 v}{\xi(\xi^2-c_+^2)}-\frac{2(c_+^2+\xi^2-2)c_+^2v^2}{(c_+^2-\xi^2)^2}+\mathcal{O}(v^3),
\end{equation}
can be solved as
\begin{align}
 v(\xi) &= \frac{c_+v_+\xi_w^2(c_+^2-\xi^2)}{\xi}\bigg/\left[c_+^3\xi+2c_+v_+\xi_w(c_+^2-2)\xi\right.\nonumber\\
 &\left.-c_+\xi_w^2(2c_+^2v_+-4v_++\xi)+ \text{Sol}\right]\label{spherical:solution}\\
 \text{Sol} &= 4v_+\xi_w^2(c_+^2-1)\left[\text{arctanh}\left(\frac{\xi}{c_+}\right)-\text{arctanh}\left(\frac{\xi_w}{c_+}\right) \right]\xi\nonumber
\end{align}
given the condition $v(\xi_w+0^+)=v_+$ at the bubble wall. Following the same procedures as in the cylindrical case, we can obtain estimate $\text{d} \ln w/ \text{d} \xi$ by first plugging~\eqref{spherical:solution} into~\eqref{eq:dw} and then expanding it to the order of $v_+^2$. Hence, $\alpha_N/ \alpha_+=w_+/w_N$ is obtained by integrating $\text{d} \ln w/ \text{d} \xi$. After inserting $v_+=\mu(\xi_w,\overline{v}_{+}(\xi_w,\alpha_+))$, $\alpha_N/ \alpha_+$ as a function of $\xi_w$ and $\alpha_+$ can be expanded in the small $\xi_w$ limit as 
\begin{align}
 \frac{\alpha_N}{\alpha_+}=1+&\frac{c_-^2-c_+^2+(1+c_-^2)\alpha_+}{2c_-^4c_+^2(1+c_+^2)}\left[c_+^2+3c_-^2+4c_+^2c_-^2\right.\nonumber\\
 &\left.-(1+c_-^2)\alpha_+\right]\xi_w^2+\mathcal{O}(\xi_w^3).
    \label{spherical:alphaLO}
\end{align}
Reversing the above relation to obtain $\alpha_{+}(\xi_w,\alpha_N)$ and substituting it into the minus-sign branch of~\eqref{eq:vbarpm}, we can derive $\overline{v}_{+}(\xi_w,\alpha_N)$ as a function of $\xi_w$ and $\alpha_N$. We next further expand $v_+=\mu(\xi_w,\overline{v}_{+}(\xi_w,\alpha_N))$ in terms of $\xi_w$ and then substitute it into~\eqref{eq:DpON}, finally the phase pressure difference can be obtained in the small $\xi_w$ limit as
\begin{align}
 \frac{p_{\rm dr}}{w_N}&=\frac{c_-^2-c_+^2+(1+c_-^2)\alpha_N}{2c_-^4(1+c_+^2)^2}\left[3c_-^2-c_+^2+2c_+^2c_-^2\right.\nonumber\\
 &\left.+(1+c_-^2)\alpha_N\right]\xi_w^2+\mathcal{O}(\xi_w^4).
 \label{spherical:pdrLO}
\end{align}
This analytical expression serves as an even better approximation in the bag limit $c_\pm\to c_s=1/\sqrt{3}$ compared to our previous estimation~\cite{Li:2023xto}, and also perfectly matches the exact numerical evaluation as shown in Fig.~\ref{fig:phase} for both cases with $c_+$ and $c_-$ although the matching is not as good as the planar and cylindrical cases as here we only include the leading-order quadratic term while the NLO quartic term is too lengthy to be informative. This leading-order pure quadratic dependence in the spherical wall velocity can be directly tested in future holographic numerical simulations of strongly coupled FOPTs with a spherical wall.

\section{Wall pressure difference}\label{sec:walldeltap}

\begin{figure*}
\centering
\includegraphics[width=0.49\textwidth]{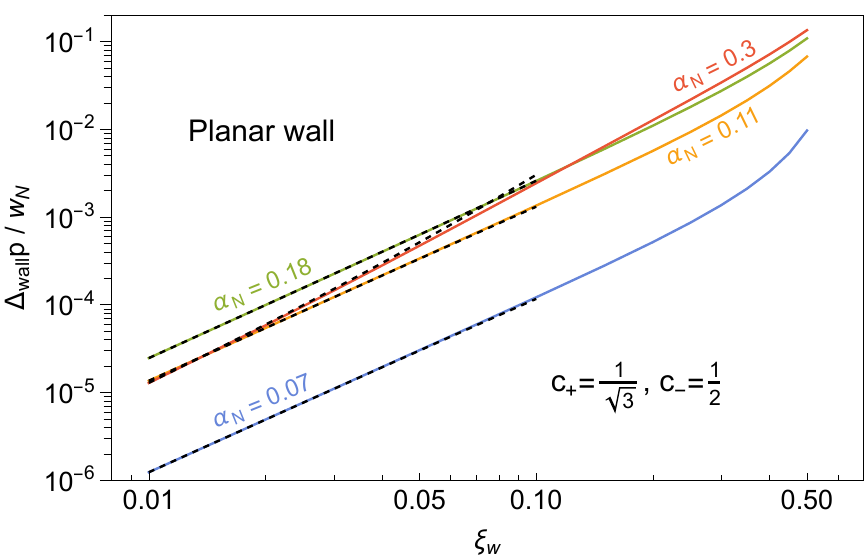}
\includegraphics[width=0.49\textwidth]{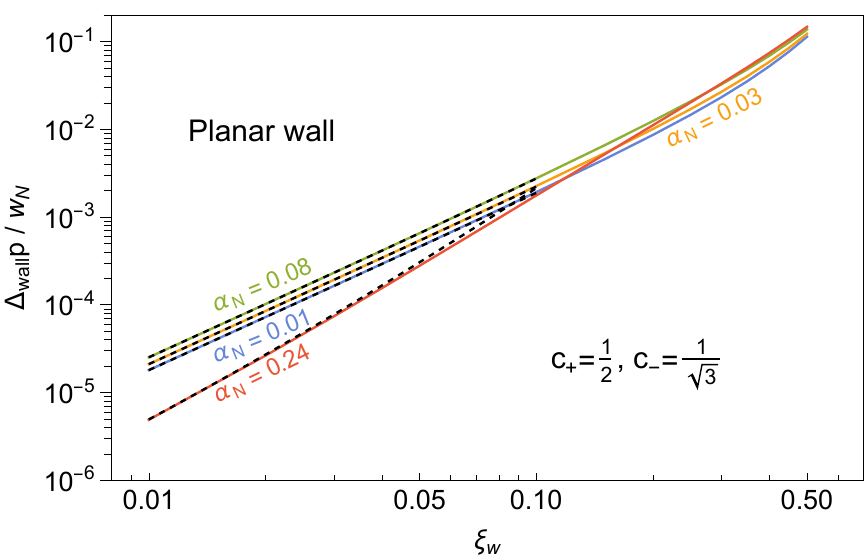}\\
\includegraphics[width=0.49\textwidth]{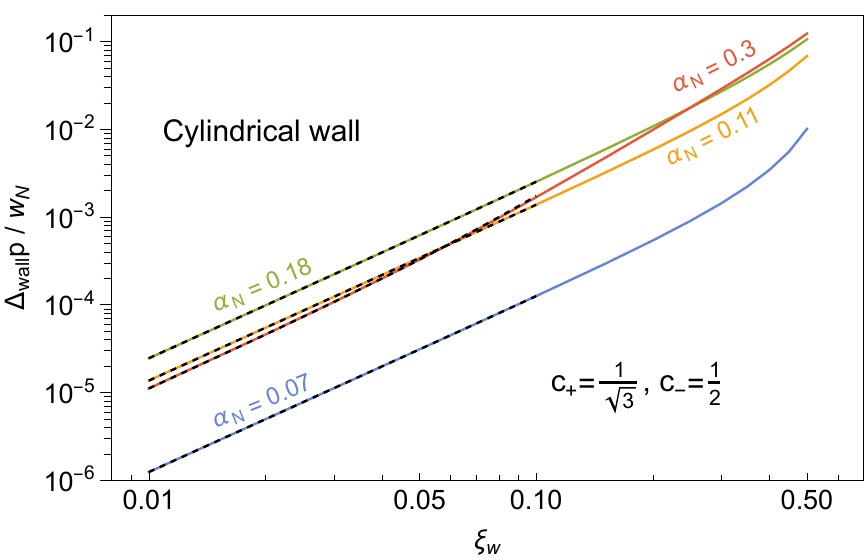}
\includegraphics[width=0.49\textwidth]{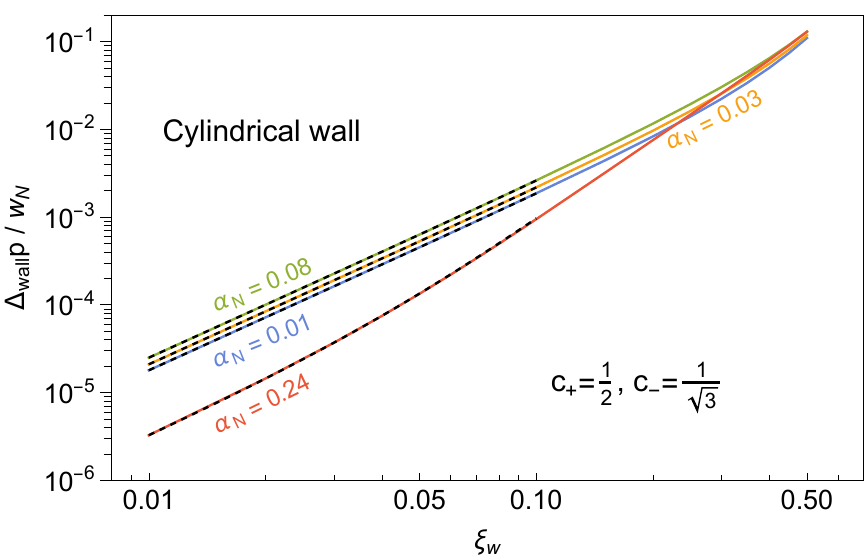}\\
\includegraphics[width=0.49\textwidth]{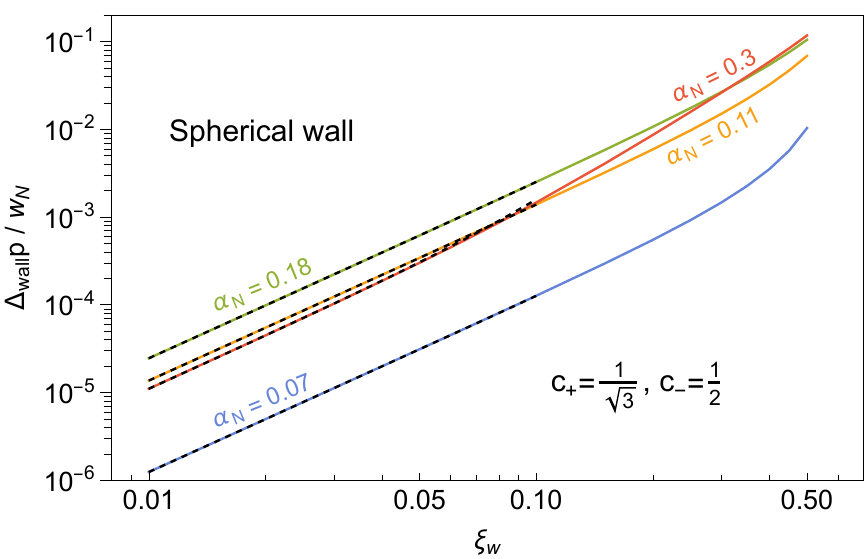}
\includegraphics[width=0.49\textwidth]{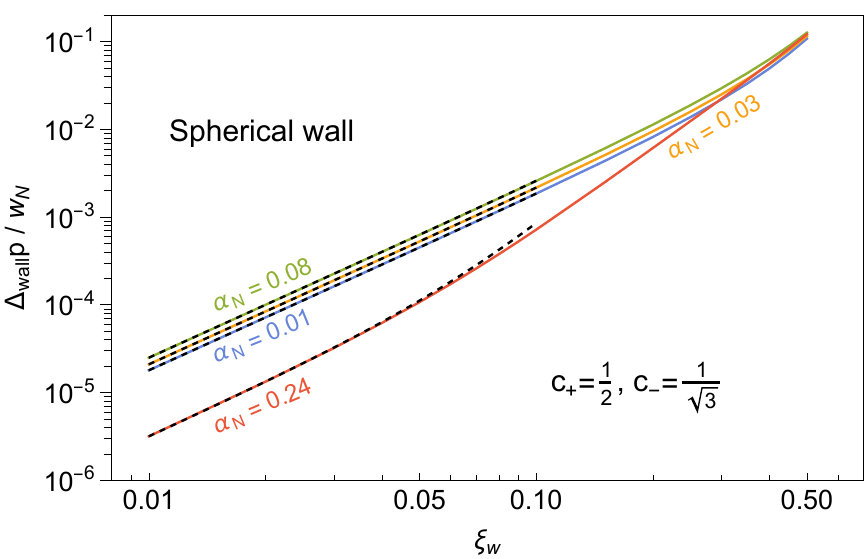}\\
\caption{The comparison between our analytical approximation~\eqref{wall:pLO} (dashed lines) [with additional next-to-leading order correction~\eqref{wall:pNLO} for relatively large $\alpha_N=0.24, 0.3$] and the exact numerical evaluations (solid lines) for the relation between the wall pressure difference  $p/w_N$ and terminal velocity $\xi_w$ (solid lines) of planar (top), cylindirical (middle), and spherical (bottom) walls, respectively, given some illustrative values for the asymptotic strength factor $\alpha_N$ in both cases with sound velocities $c_+>c_-$ (left) and $c_+<c_-$ (right).}
\label{fig:wall}
\end{figure*}

Apart from the phase pressure difference away from the bubble wall, we can also approximate in the nonrelativistic limit for the pressure difference near the bubble wall, $\Delta_\mathrm{wall}p\equiv p_+-p_-$, which can evaluated by the junction condition~\eqref{eq:WallJunction2},
\begin{equation}
\label{eq:WallDpJunction}
    \frac{\Delta_{\rm wall}p}{w_N}=\frac{\overline{\gamma}_-^2\overline{v}_-^2w_--\overline{\gamma}_+^2\overline{v}_+^2w_+}{w_N}.
\end{equation}
For the deflagration mode with $\overline{v}_-=\xi_w$ and $\overline{v}_+=\mu(\xi_w,v_+)$, we can solve $\alpha_+$ from the minus-branch of~\eqref{eq:vbarpm} as
\begin{align}
    \label{alpha:deflagration}
    \alpha_+&=\frac{1}{(c_-^2+1)(v_+^2-1)\xi_w}\left[v_+\xi_w(\xi_w-v_+)+c_+^2\xi_w(v_+\xi_w-1)\right.\nonumber\\
    &\left.+c_-^2(\xi_w-v_+-v_+c_+^2+v_+^2c_+^2\xi_w)\right].
\end{align}
Combining~\eqref{eq:WallDpJunction} and~\eqref{alpha:deflagration}, we can obtain the wall pressure difference $\Delta_{\rm wall}p/w_N$ in terms of $c_\pm$, $\xi_w$, $\alpha_N$, and $v_+$ as
\begin{align}
    \label{wall:pressure}
    &\frac{\Delta_{\rm wall}p}{w_N}
    =\left[(1+c_-^2)(v_+-\xi_w)\xi_w v_+\alpha_N\right]\bigg/\left\{\xi_w v_+(\xi_w-v_+)\right.\nonumber\\
    &\left.+c_+^2\xi_w(\xi_w v_+-1)+c_-^2[\xi_w+v_+(c_+^2\xi_w v_+-c_+^2-1)]\right\},
\end{align}
where we have converted $w_-$ to $w_+$ via the junction condition~\eqref{eq:WallJunction1}, and then converted $w_+$ to $w_N$ via $w_+\alpha_+=w_N\alpha_N$. After plugging the nonrelativistic analytic approximations $v_+(\xi_w,\alpha_N)$ we obtained in the previous three subsections for planar, cylindrical, and spherical walls into~\eqref{wall:pressure}, we finally arrive at a universal quadratic dependence in the wall velocity at the leading order for the wall pressure difference in the small $\xi_w$ limit as
\begin{equation}
\label{wall:pLO}
    \left(\frac{\Delta_{\rm wall}p}{w_N}\right)_{\rm LO}^{D=0,1,2}=\frac{(1+c_-^2)(c_+^2-\alpha_N)[c_-^2-c_+^2+\alpha_N(1+c_-^2)]}{c_-^4(1+c_+^2)^2}\xi_w^2,
\end{equation}
whose bag limit $c_\pm\to c_s$ is the same as our previous result~\cite{Li:2023xto},
\begin{align}
\left(\frac{p_+-p_-}{w_N}\right)_{\mathrm{LO}, c_\pm\to c_s}^{D=0,1,2}=\left(\frac{\alpha_N}{c_s^2}-\frac{\alpha_N^2}{c_s^4}\right)\xi_w^2.
\end{align}
This universal scaling for different wall shapes can be understood as the pressure difference taken near the wall does not care about its global shape. This is different from the phase pressure difference taken between the null infinity and bubble center, which does care about the global shapes of the bubble wall, containing not only the information near the bubble wall but also the whole bubble-fluid system including the sound shell and shock-wave front (if any). This is why the phase pressure difference admits different leading-order behaviors, that is, the leading-order linear, logarithmic-quadratic, and purely quadratic dependences for the planar, cylindrical, and spherical walls, respectively. Nevertheless, for the asymptotic strength factor $\alpha_N$ taking a relatively large value, the leading-order analytical approximation is not enough, and we must consider the next leading-order correction,
\begin{widetext}
\begin{equation}
\begin{split}
    \label{wall:pNLO}
    \left(\frac{p_+-p_-}{w_N}\right)_{\rm NLO}^{D=0}&=\frac{(1+c_-^2)[c_-^2-c_+^2+\alpha_N(1+c_-^2)][c_+^2c_-^2-c_+^4+\alpha_N^2(1+c_-^2)]}{c_-^6c_+(1+c_+^2)^2}\xi_w^3,\\
    \left(\frac{p_+-p_-}{w_N}\right)_{\rm NLO}^{D=1}&=\frac{(1+c_-^2)[c_-^2-c_+^2+\alpha_N(1+c_-^2)]}{2c_-^8c_+^2(1+c_+^2)^4}\left(\text{Cy}_1+\text{Cy}_2+\text{Cy}_3+\text{Cy}_4\right)\xi_w^4,     \\
    \left(\frac{p_+-p_-}{w_N}\right)_{\rm NLO}^{D=2}&=\frac{(1+c_-^2)[c_-^2-c_+^2+\alpha_N(1+c_-^2)]}{2c_-^8c_+^2(1+c_+^2)^4}\left(\text{Sp}_1+\text{Sp}_2+\text{Sp}_3\right)\xi_w^4,\\
\end{split}
\end{equation}
\end{widetext}
with
\begin{widetext}
\begin{equation}
    \begin{split}
    \text{Cy}_1&=(c_+^2-\alpha_N)^2(c_+^4+3c_+^2+c_+^2\alpha_N-\alpha_N)+(4\ln 2) c_-^2 c_+^2\alpha_N^2+c_-^2\alpha_N(c_+^2-\alpha_N)(1+2\alpha_N-2\ln2),                            \\
    \text{Cy}_2&=-c_-^2(c_+^2-\alpha_N)[(2\ln 2)c_+^6-c_+^4[4+3\alpha_N-(2\ln2)\alpha_N]+c_+^2(2\alpha_N^2+4\alpha_N+2\ln2)],                           \\
    \text{Cy}_3&=c_-^4\left[c_+^6(2+2\ln2+2\alpha_N)+c_+^4[(2\ln2-4)\alpha_N^2-5\alpha_N-1+4\ln2]+\text{Cy}_3^{'}\right],                            \\
    \text{Cy}_3^{'}&=c_+^2[\alpha_N^3+(4\ln 2+1)\alpha_N^2-\alpha_N+2\ln2-1]-\alpha_N^2(1-2\ln 2+\alpha_N)                        \\
    \text{Cy}_4&=2c_-^2\left[(\alpha_N^2(1+c_+^2)^2+c_-^2(1+c_+^2)^2(c_+^2+\alpha_N^2)-(1+c_+^4)c_+^4)\ln\left(\frac{c_+}{\xi_w}\right)-2c_+^6\ln\left(\frac{2c_+}{\xi_w}\right)\right]                            \\
    \text{Sp}_1&=(c_+^2-\alpha_N)^2(c_+^4+3c_+^2+c_+^2\alpha_N-\alpha_N),\\
     \text{Sp}_2&=-c_-^2(c_+^2-\alpha_N),[4c_+^6+c_+^4(4+\alpha_N)+2c_+^2(\alpha_N^2+6\alpha_N+2)+\alpha_N(3-2\alpha_N)],\\
    \text{Sp}_3&=c_-^4[2c_+^6(3+\alpha_N)+c_+^4(7-5\alpha_N)+c_+^2(\alpha_N^3+9\alpha_N^2-\alpha_N+3)+(3-\alpha_N)\alpha_N^2].
    \end{split}
\end{equation}
\end{widetext}
The comparison between our analytical approximation~\eqref{wall:pLO} [with additional next-to-leading order correction~\eqref{wall:pNLO} for a relatively large $\alpha_N=0.24, 0.3$] and the exact numerical evaluations is presented in Fig.~\ref{fig:wall} with perfect match in the nonrelativistic limit.
Note that the crossing of curves for relatively large $\alpha_N$ at relatively large $\xi_w$ is due to the nonmonotonous dependence of the wall pressure difference on $\alpha_N$ at relatively large $\xi_w$. This can be easily illustrated in the case of a simple bag EOS~\cite{Li:2023xto} with $c_+=c_-=1/\sqrt{3}$, in which case the wall pressure difference $p_+-p_-$ reads
\begin{equation}
    \label{eq:WallDifference}
    \frac{p_+-p_-}{w_N}=\frac{w_+}{w_N}\overline{\gamma}_{+}^{2}\overline{v}_+(\overline{v}_--\overline{v}_+),
\end{equation}
after using the junction conditions $w_-\overline{\gamma}_-^2\overline{v}_-=w_+\overline{\gamma}_+^2\overline{v}_+$ and $w_-\overline{\gamma}^2_-\overline{v}_-^2+p_-=w_+\overline{\gamma}^2_+\overline{v}_+^2+p_+$. When the bubble wall velocity $\xi_w$ is small, the fluid profile is deflagration and hence the wall-frame fluid velocity just behind the wall reads $\overline{v}_-=\xi_w$. For $w_+/w_N=\alpha_N/\alpha_N=1+\mathcal{O}(\xi_w)$, we take $w_+/w_N \simeq 1$ and then \eqref{eq:WallDifference} turns into
\begin{equation}
    \label{eq:walldifference}
    \frac{p_+-p_-}{w_N}=\frac{\overline{v}_+}{1-\overline{v}_+^{2}}(\xi_w-\overline{v}_+).
\end{equation}
As one can explicitly check numerically, although the wall-frame fluid velocity just in front of the wall $\bar{v}_+$ decreases with an increasing $\alpha_N$, the wall-pressure difference is not a monotonic function of $\bar{v}_+$, and hence it is also nonmonotonic to $\alpha_N$. For example, for a small $\xi_w$, the leading-order wall pressure difference in the bag case,
\begin{equation}
    \label{eq:LObag}
    \frac{\Delta_{\rm wall}p}{w_N}\bigg|_{\rm LO}=9\left(\frac{1}{3}-\alpha_N\right)\alpha_N \xi_w^2,
\end{equation}
will be larger if $\alpha_N$ is closer to $1/6$. However, when $\xi_w$ is relatively larger,  we need take into account the NLO term, 
\begin{equation}
    \label{eq:NLObag}
    \frac{\Delta_{\rm wall}p}{w_N}\bigg|_{\rm NLO}=9(\alpha_N-8\alpha_N^2+36\alpha_N^3-9\alpha_N^4)\xi_w^4.
\end{equation}
where the quartic coefficient increases as $\alpha_N$ increases. Therefore, when $\xi_w$ is relatively large, the wall pressure difference of large $\alpha_N$ is larger than that of small $\alpha_N$.

\section{Conclusions and discussions}\label{sec:condis}

\begin{figure*}
    \centering
    \includegraphics[width=0.52\textwidth]{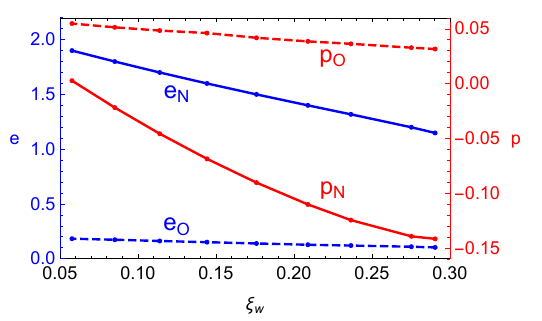}
    \includegraphics[width=0.46\textwidth]{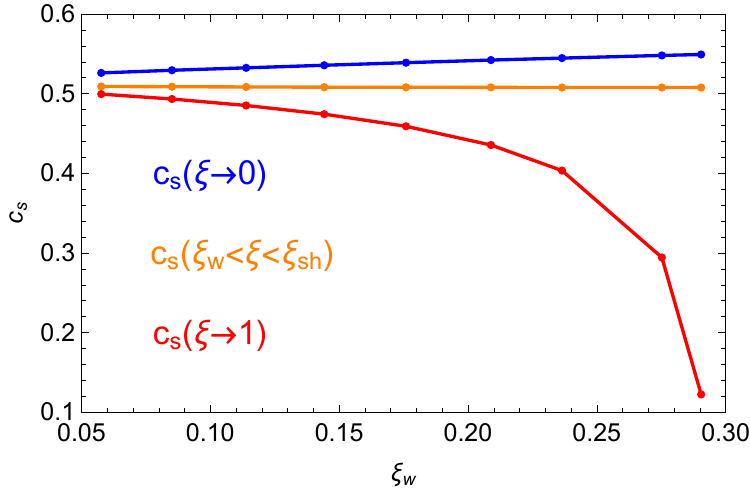}\\
    \includegraphics[width=0.49\textwidth]{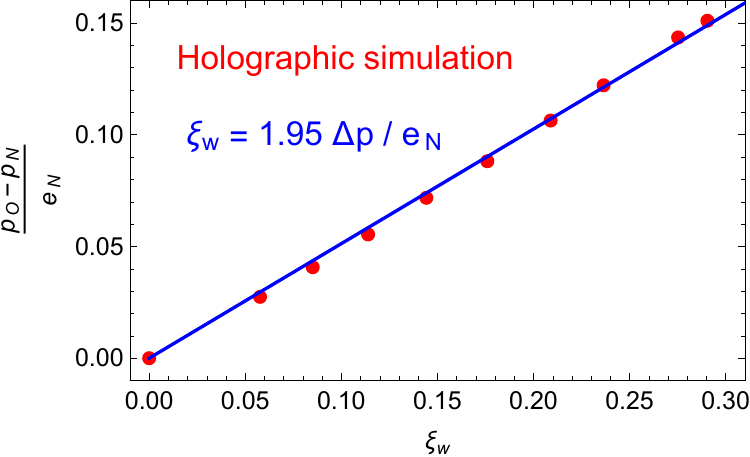}
    \includegraphics[width=0.49\textwidth]{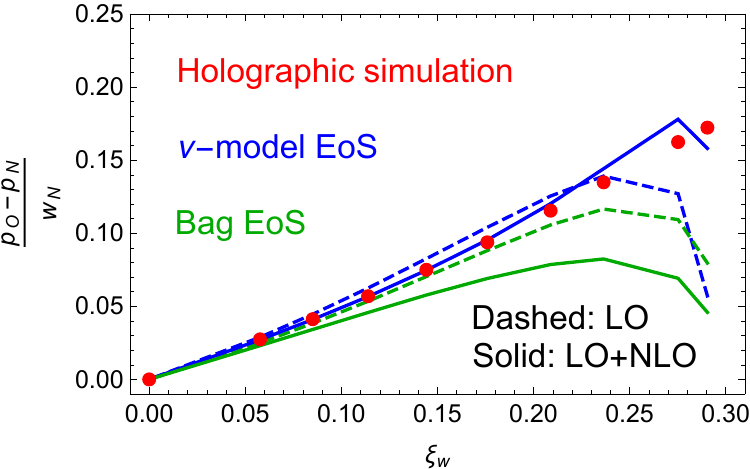}\\
    \caption{The original data points from Ref.~\cite{Bea:2021zsu} for the energy density and pressure (first panel) as well as sound velocity (second panel) with respect to the wall velocity. The third panel reproduces their original fit to the phase pressure difference in unit of asymptotic energy density, which is actually achieved highly nontrivial by adjusting the strength factor and EOS simultaneously. The last panel compares the data points from the holographic numerical simulation to our analytic approximation~\eqref{planar:pdrLO} from both bag EOS and $\nu$-model EOS.}
    \label{fig:FitPlanar}
\end{figure*}

The cosmological FOPT serves as an indispensable probe into the early Universe for the new physics beyond the standard model of particle physics. The weakly coupled FOPT is widely studied for its validity in adopting the perturbative field theory method to estimate the vacuum decay rate and bubble wall velocity. However, this is not the case for the strongly coupled FOPTs where the perturbative method ceases to apply for lack of perturbative definitions on the effective potential and collision terms in Boltzmann equations. Fortunately, the holographic method as a specific realization of the strong-weak duality can map the strongly coupled FOPT on the boundary into a weakly coupled gravity theory in the bulk. Recent holographic numerical simulations of strongly coupled FOPTs not only prefer a nonrelativistic terminal wall velocity but also confirm the perfect-fluid hydrodynamics approximation, and in particular, reveal an intriguing linear correlation between the phase pressure difference and terminal velocity of planar wall. By fully appreciating the perfect-fluid hydrodynamics, we analytically reproduce such a correlation not only for the planar wall but also for cylindrical and spherical walls in the case with a bag EOS. To be more close to the realistic case, we generalize in this paper our previous analytic results into the case with a $\nu$-model EOS beyond the simple bag model. The analytic approximations of the phase pressure difference~\eqref{planar:pdrLO},~\eqref{cylindrical:pdrLO}, and~\eqref{spherical:pdrLO} we obtained for the planar, cylindrical, and spherical walls, respectively, not only well-match the exact numerical evaluations from the perfect-fluid hydrodynamics, but also improve our previous results in the bag limit $c_\pm\to c_s$,
\begin{align}
\left(\frac{p_O-p_N}{w_N}\right)_{c_\pm\to c_s}^{D=0}&=\frac{\alpha_N}{c_s}\xi_w+\mathcal{O}(\xi_w^2),\\
\left(\frac{p_O-p_N}{w_N}\right)_{c_\pm\to c_s}^{D=1}&=\left[\frac{\alpha_N^2}{2c_s^4}-\frac{\alpha_N}{c_s^2}\left(1+\ln\frac{\xi_w}{2c_s}\right)\right]\xi_w^2+\mathcal{O}(\xi_w^4),\\
\left(\frac{p_O-p_N}{w_N}\right)_{c_\pm\to c_s}^{D=2}&=\left(\frac{\alpha_N}{c_s^2}+\frac{\alpha_N^2}{2c_s^4}\right)\xi_w^2+\mathcal{O}(\xi_w^4).
\end{align}
All these analytic results can be directly tested in future holographic numerical simulations (see, for example, the last panel of Fig.~\ref{fig:FitPlanar} for a  perfect match between our analytic approximation~\eqref{planar:pdrLO} and holographic numerical simulation~\cite{Bea:2021zsu} in the case with an expanding planar wall), which would shed light on the understanding of strongly coupled FOPT and its holographic dual.

\begin{acknowledgments}
We thank Mikel Sanchez Garitaonandia for helpful correspondence and for providing the original data. We also thank an anonymous referee for insightful suggestions in greatly improving the presentations.
S.J.W. is supported by the National Key Research and Development Program of China Grant  No. 2021YFC2203004, No. 2020YFC2201501 and No. 2021YFA0718304, 
the National Natural Science Foundation of China Grants 
No. 12105344, No. 12235019, and No. 12047503,
the Key Research Program of the Chinese Academy of Sciences (CAS) Grant No. XDPB15, 
the Key Research Program of Frontier Sciences of CAS, 
and the Science Research Grants from the China Manned Space Project No. CMS-CSST-2021-B01.
\end{acknowledgments}

\appendix

\section{Hydrodynamics beyond bag EOS}\label{app:numodel}

In this appendix, we revisit in detail the hydrodynamics beyond the bag EOS specifically in the $\nu$-model~\cite{Leitao:2014pda} where the sound velocity profile $c_s(\xi)$ takes constant values $c_-$ and $c_+$ inside ($\xi<\xi_w$) and outside ($\xi>\xi_w$) of the bubble wall, respectively. This $\nu$-model EOS together with the junction conditions~\eqref{eq:WallJunction1} and~\eqref{eq:WallJunction2} across the bubble wall gives rise to the hydrodynamic solution~\eqref{eq:vbarpm} which we repeat here at your convenience,
\begin{equation}
\bar{v}_{+}=\frac{1}{1+\alpha_{+}}\left( qX_{+}\pm \sqrt{q^2X_{-}^2+\alpha_{+}^2+(1-c_{+}^2)\alpha_+ +q^2c_{-}^2-c_{+}^2} \right).\label{eq:appvbarpm}
\end{equation}
Here abbreviations $q\equiv(1+c_{+}^2)/(1+c_{-}^2)$, $X_{\pm}\equiv \bar{v}_{-}/2\pm c_{-}^2/(2\bar{v}_{-})$, $\alpha_{+}\equiv \Delta V_0 /(a_+ T_+^{\nu_+})$, and $\Delta V_0 \equiv V_0^{+}-V_0^{-}$ are introduced for clarity. Similar to the bag-EOS case, the detonation (deflagration) mode picks the plus-sign (minus-sign) branch of~\eqref{eq:appvbarpm}. Note that in order for $\bar{v}_+$ in~\eqref{eq:appvbarpm} to be real positive number, $\alpha_+$ should be bounded from below by $(c_+^2-c_-^2)/(1+c_-^2)$, and for $\alpha_+>c_+^2$, only the detonation mode exists. Note also that the condition $\bar{v}_-=c_-$ defines the Jouguet velocity, which further defines the Jouguet detonation (deflagration) mode when $\bar{v}_+$ in the plus-sign (minus-sign) branch of~\eqref{eq:appvbarpm} takes its minimal (maximal) value as 
\begin{align}
v_{J}^\mathrm{det}(\alpha_+)&=\frac{qc_-+\sqrt{q^2c_-^2-(1+\alpha_+)(c_+^2-\alpha_+)}}{1+\alpha_+},\label{Jouguet:detonation}\\
v_{J}^\mathrm{def}(\alpha_+)&=\frac{qc_--\sqrt{q^2c_-^2-(1+\alpha_+)(c_+^2-\alpha_+)}}{1+\alpha_+}.\label{Jouguet:deflagration}
\end{align}
It is worth noting that if $\alpha_+$ takes its minimal value $(c_+^2-c_-^2)/(1+c_-^2)$, we have $v^{\rm det}_{J}|_{\rm min}=v^{\rm def}_{J}|_{\rm max}=c_-$, that is to say, we always have $v^{\rm det}_{J}\geq c_-$ and $v^{\rm def}_{J}\leq c_-$.

\begin{figure*}
\centering
\includegraphics[width=0.49\textwidth]{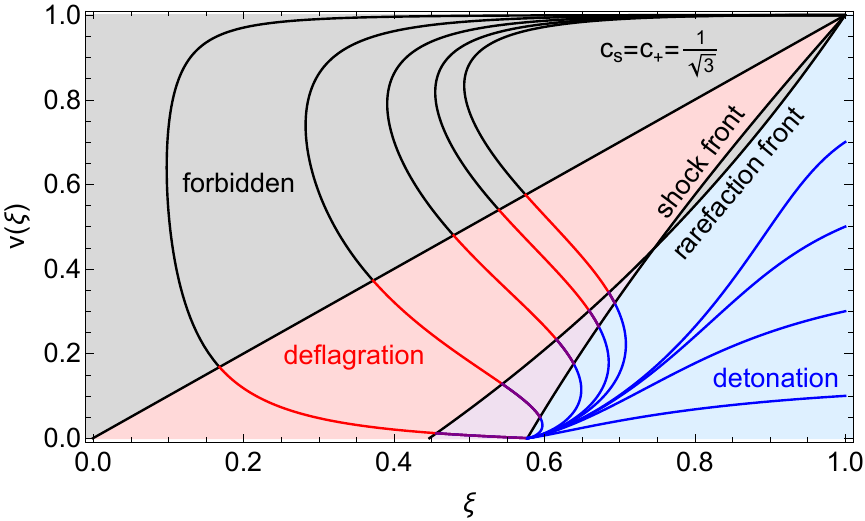}
\includegraphics[width=0.49\textwidth]{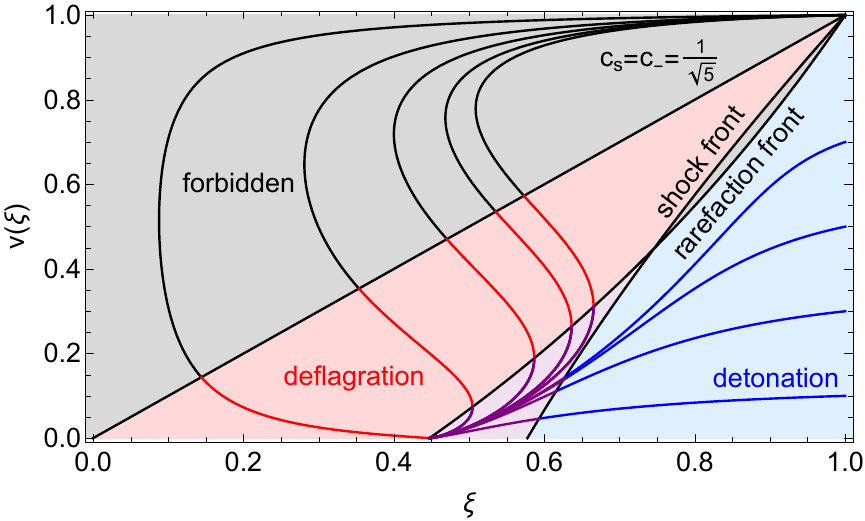}\\
\caption{The fluid velocity profiles $v(\xi)$ for a $\nu$-model EOS with $c_s=c_+=1/\sqrt{3}$ (left) and $c_s=c_-=1/\sqrt{5}$ (right). The shockwave front is defined by $\mu(\xi,v)\xi=c_+^2$ while the rarefaction front is defined by $\mu(\xi,v)=c_-$. The gray, red, blue, and purple shaded regions correspond to the forbidden, deflagration, detonation, hybrid profiles, respectively.}
\label{fig:VelocityProfile}
\end{figure*}

After specifying the physical branches of hydrodynamic solutions for different expansion modes, we can solve the fluid velocity profile $v(\xi)$ from the hydrodynamic EoM~\eqref{eq:dv} given corresponding junction conditions~\eqref{eq:WallJunction1} and~\eqref{eq:ShockJunction1} at the bubble wall and shockwave front, if any. We illustrate the fluid velocity profiles $v(\xi)$ naively solved from~\eqref{eq:dv} in Fig.~\ref{fig:VelocityProfile} for some particular values of the sound velocity.  Note that $(\xi=c_s,v=0)$ is an improper node of~\eqref{eq:dv}, where all of $v(\xi)$ curves are approached from $c_s+0^+$. The expansion modes are separated by the rarefaction front $\mu(\xi,v)=c_-$ and shockwave front $\mu(\xi,v)\xi=c_+^2$. We next turn to solve the fluid velocity and enthalpy profiles specifically for different expansion modes.

\subsection{Weak detonation}\label{app:WeakDetona}

\begin{figure*}
\centering
\includegraphics[width=0.49\textwidth]{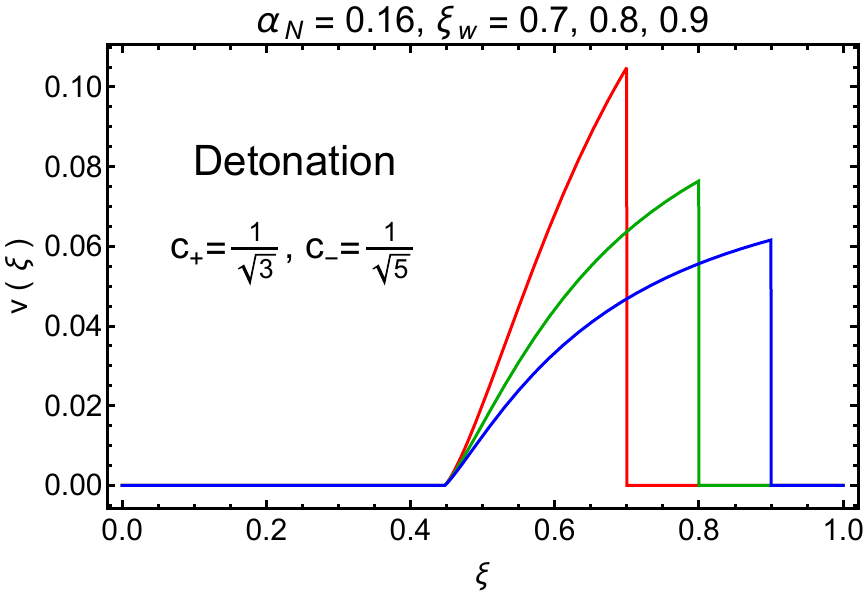}
\includegraphics[width=0.49\textwidth]{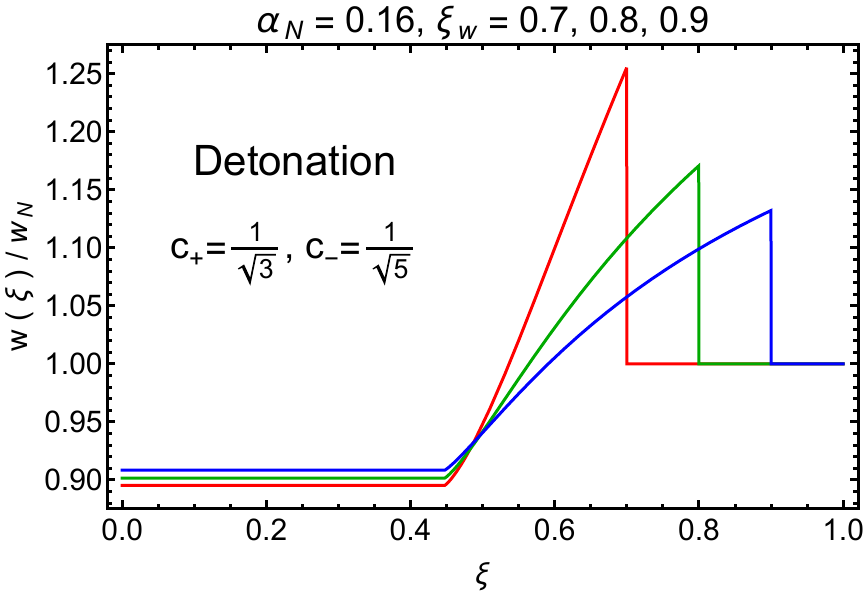}\\
\caption{The profiles of the fluid velocity $v(\xi)$  (left) and enthalpy $w(\xi)/w_N$ (right) for the weak detonation mode. We fix $\alpha_N=0.16$ and take $c_+=1/\sqrt{3}$, $c_-=1/\sqrt{5}$ for illustration. The red, green, and blue curves correspond to the bubble wall velocities $\xi_w=0.7, 0.8, 0.9$, respectively.}
\label{fig:WeakDetona}
\end{figure*}

The detonation mode is defined when the fluid velocity in front of the bubble wall is vanished, $v(\xi>\xi_w)=0$, namely $\bar{v}_+=\xi_w$ in the wall frame. Thus, $\bar{v}_-$ can be solved from the plus-sign branch of~\eqref{eq:appvbarpm}, leading directly to $v_-=\mu(\xi_w,\bar{v}_-)$. Hence the condition $v_->v_+=0$ namely $\bar{v}_+>\bar{v}_-$ defines the detonation mode. The detonation mode can be of either weak or Jouguet types with $\bar{v}_->c_-$ or $\bar{v}_-=c_-$, that is $\xi_w>v_J^\mathrm{det}$ or $\xi_w=v_J^\mathrm{det}$, respectively. We postpone the discussion of the Jouguet detonation until Sec.~\ref{app:JouguetDetona}, but first solve here the hydrodynamic EoM~\eqref{eq:dv} with $c_s=c_-$ for the fluid velocity profile $v(\xi)$ passing through $(\xi_w, v_-)$ in the case of weak detonation ($\xi_w>v_J^\mathrm{det}$) as illustrated in the left panel of Fig.~\ref{fig:WeakDetona}. Note that for the $\nu$-model EOS, the weak detonation mode contains not only the case with a large $\xi_w>v_J^\mathrm{det}>c_+$ but also the case with a very large $\xi_w>c_+^2/v_J^\mathrm{det}>v_J^\mathrm{det}$. As a comparison for a bag EOS with $c_+=c_-=c_s$, only the former case $\xi_w>v_J^\mathrm{det}>c_s$ survives. With the fluid velocity profile $v(\xi)$ solved from~\eqref{eq:dv} at hand, the corresponding enthalpy profile $w(\xi)$ can be obtained simply by integrating~\eqref{eq:dw} from the point $(\xi_w,w_-)$ with the enthalpy $w_-$ just behind the wall determined by the junction condition~\eqref{eq:WallJunction1} from the enthalpy $w_+=w_N$ in front of the wall up to the null infinity. We illustrate the enthalpy profile for the weak detonation in the right panel of Fig.~\ref{fig:WeakDetona}.

\subsection{Weak deflagration}\label{app:WeakDeflag}

\begin{figure*}
\centering
\includegraphics[width=0.49\textwidth]{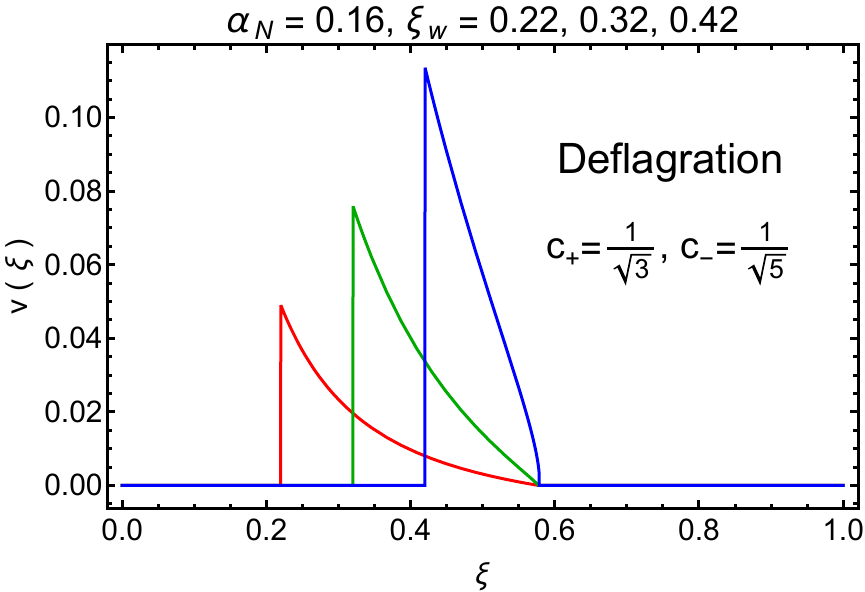}
\includegraphics[width=0.49\textwidth]{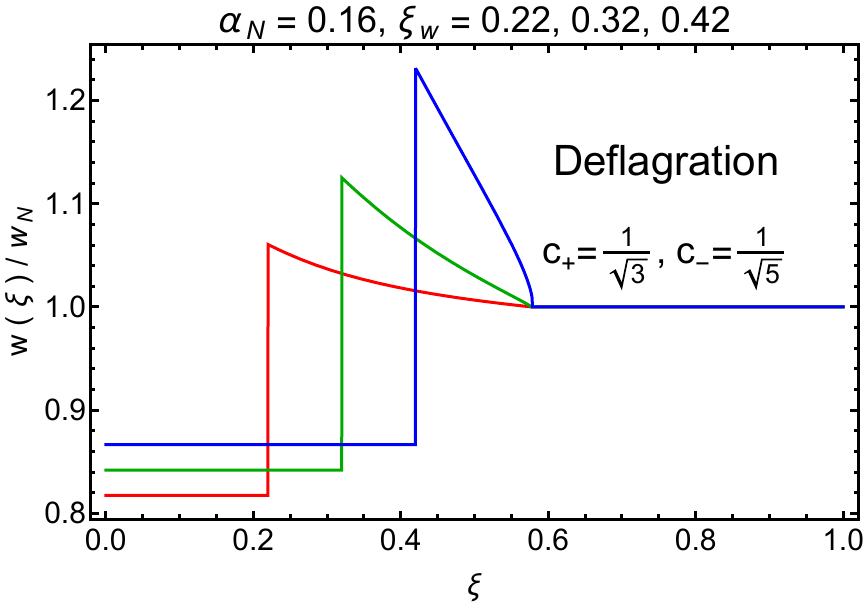}\\
\caption{The fluid profiles $v(\xi)$ (left) and $w(\xi)/w_N$ (right) of the weak deflagration mode. We fix $\alpha_N=0.16$ and take $c_+=1/\sqrt{3}$, $c_-=1/\sqrt{5}$. The red, green, and blue lines correspond to the bubble wall velocity $\xi_w$ takes the value of 0.22, 0.32 and 0.42, respectively.}
\label{fig:WeakDeflag}
\end{figure*}

The deflagration mode is defined when the fluid velocity behind the bubble wall is vanished, $v(\xi<\xi_w)=0$, namely $\bar{v}_-=\xi_w$ in the wall frame. Thus, $\bar{v}_+$ can be solved from the minus-sign branch of~\eqref{eq:appvbarpm}, leading directly to $v_+=\mu(\xi_w,\bar{v}_+)$. Hence the condition $v_+>v_-=0$ namely $\bar{v}_->\bar{v}_+$ defines the deflagration mode. The deflagration mode can be of either weak or Jouguet types with $\xi_w=\bar{v}_-<c_-$ or $\xi_w=\bar{v}_-=c_-$, that is $\xi_w<v_J^\mathrm{def}$ or $\xi_w=v_J^\mathrm{def}$, respectively. We postpone the discussion of the Jouguet deflagration until Sec.~\ref{app:JouguetDeflag}, but first solve here the hydrodynamic EoM~\eqref{eq:dv} with $c_s=c_-$ for the fluid velocity profile $v(\xi)$ passing through $(\xi_w,v_+)$ in the case of weak deflagration ($\xi_w<v_J^\mathrm{def}$) as illustrated in the left panel of Fig.~\ref{fig:WeakDeflag}. Note that the solved fluid velocity profile $v(\xi)$ should be cut off due to the shockwave front at $\xi_{sh}$ with corresponding fluid velocity $v_{sh}\equiv v(\xi_{sh}+0^-)$, both of which can be determined as shown shortly below. First, it is easy to find $\tilde{v}_L\tilde{v}_R=c_+^2$ for the shock-frame fluid velocities $\tilde{v}_{L/R}$ just inside/outside of the shockwave front since the whole shockwave is in the symmetric phase in front of the bubble wall. Then, as the fluid velocity in front of the shockwave front is at rest, $v_R=\mu(\xi_{sh},\tilde{v}_R)=0$, the shockwave front velocity $\xi_{sh}=\tilde{v}_R=c_+^2/\tilde{v}_L=c_+^2/\mu(\xi_{sh},v_L)$ can be directly solved from $v_L=v(\xi_{sh}+0^-)\equiv v_{sh}$ given by extrapolating the solved profile of $v(\xi)$ from $(\xi_w, v_+)$ to $(\xi_{sh},v_{sh})$. The enthalpy profile $w(\xi)$ shown in the right panel of Fig.~\ref{fig:WeakDeflag} can be obtained by integrating the fluid EoM~\eqref{eq:dw} from the shock front $(\xi_{sh}, w_L)$ all the way back to the wall, where $w_L\equiv w(\xi_{sh}+0^-)$ can be determined by the junction condition~\eqref{eq:ShockJunction1} with $w_R=w_N$, $\tilde{v}_R=\xi_{sh}$, and $\tilde{v}_L=\mu(\xi_{sh},v_{sh})$. At the bubble wall, the enthalpy profile experiences a sudden jump from $w_+=w(\xi_w+0^+)$ to $w_-$ determined by the junction condition~\eqref{eq:WallJunction1} with $\bar{v}_-=\xi_w$ and $\bar{v}_+(\alpha_+,\xi_w)$ given by the minus-sign branch of~\eqref{eq:vbarpm}.

\subsection{Jouguet deflagration}\label{app:JouguetDeflag}

\begin{figure*}
\centering
\includegraphics[width=0.49\textwidth]{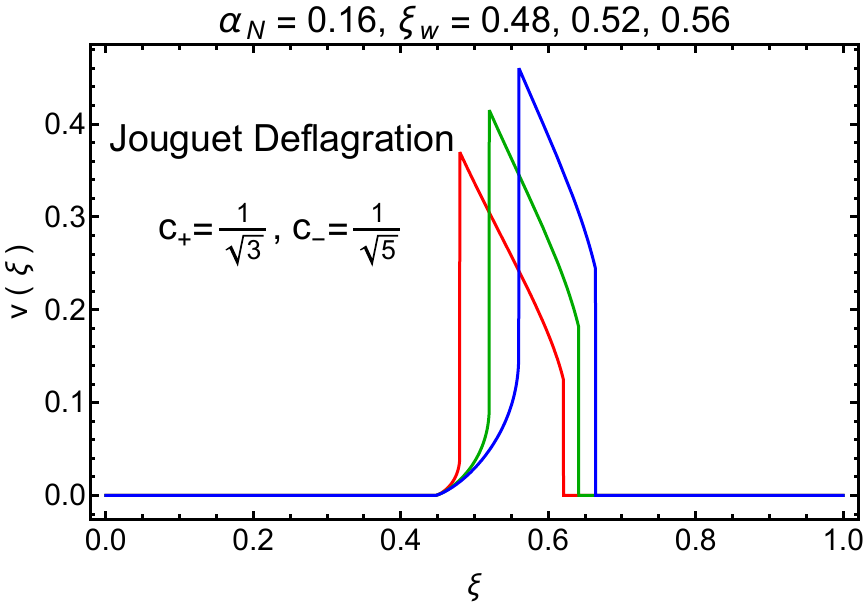}
\includegraphics[width=0.49\textwidth]{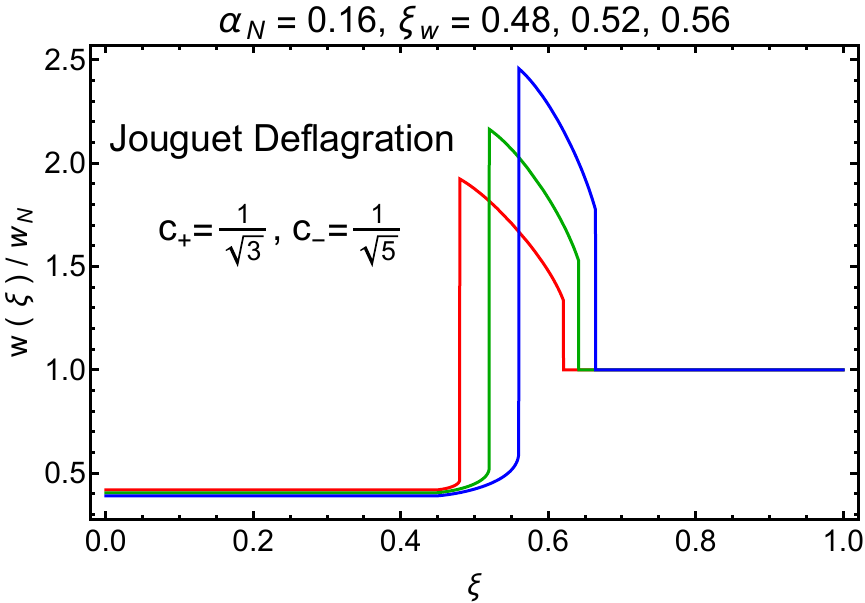}\\
\caption{The fluid profiles $v(\xi)$ (left) and $w(\xi)/w_N$ (right) of the Jouguet deflagration mode. We fix $\alpha_N=0.16$ and take $c_+=1/\sqrt{3}$, $c_-=1/\sqrt{5}$. The red, green and blue lines correspond to the bubble wall velocity $\xi_w$ taking the value of 0.48, 0.52, and 0.56, respectively.}
\label{fig:JouguetDeflag}
\end{figure*}

The Jouguet deflagration mode (or we call it the hybrid mode in the bag model) is a special deflagration mode  ($\overline{v}_+<\overline{v}_-$) of Jouguet type ($\overline{v}_-=c_-$) corresponding to the minus-sign branch of~\eqref{eq:vbarpm} and realized with the wall velocity lying between $c_-<\xi_w<v_{J}^{\rm det}$. The fluid velocity profile in Fig.~\ref{fig:JouguetDeflag} contains both compressive shockwave and rarefaction wave in the front and back of the bubble wall, respectively, as derived shortly below. The Jouguet deflagration condition $\overline{v}_{-}=c_-$ leads to $\overline{v}_+=v_{J}^{\rm def}$ from~\eqref{Jouguet:deflagration} by~\eqref{eq:vbarpm}, giving rise to $v(\xi_w+0^+)\equiv v_+=\mu(\xi_w,v_J^\mathrm{def})$ and $v(\xi_w+0^-)\equiv v_-=\mu(\xi_w,c_-)$ that can be used to solve the fluid EoM~\eqref{eq:dv} both forward and backward from $(\xi_w+0^+,v_+)$ and $(\xi_w+0^-,v_-)$ with $c_s=c_+$ and $c_s=c_-$, respectively. The solved velocity profile again vanishes in front of the shockwave front $\xi=\xi_{sh}+0^+$ and behind $\xi=c_-+0^-$ as in the weak deflagration and weak detonation cases. The enthalpy profile can be similarly obtained from integrating~\eqref{eq:dw} backward from both $w(\xi_{sh}+0^-)=w_L$ and $w(\xi_w+0^-)=w_-$, where the enthalpies $w_L$ and $w_-$ are sequentially determined by the junction conditions~\eqref{eq:ShockJunction1} and~\eqref{eq:WallJunction1} with $w_R=w_N$, $\tilde{v}_R=\xi_{sh}$, $\tilde{v}_L=\mu(\xi_{sh},v_{sh})$ and $w_+=w(\xi_w+0^+)$, $\overline{v}_+=v_{J}^{\rm def}$, $\overline{v}_-=c_-$, respectively.

\subsection{Jouguet detonation}\label{app:JouguetDetona}

\begin{figure*}
\centering
\includegraphics[width=0.49\textwidth]{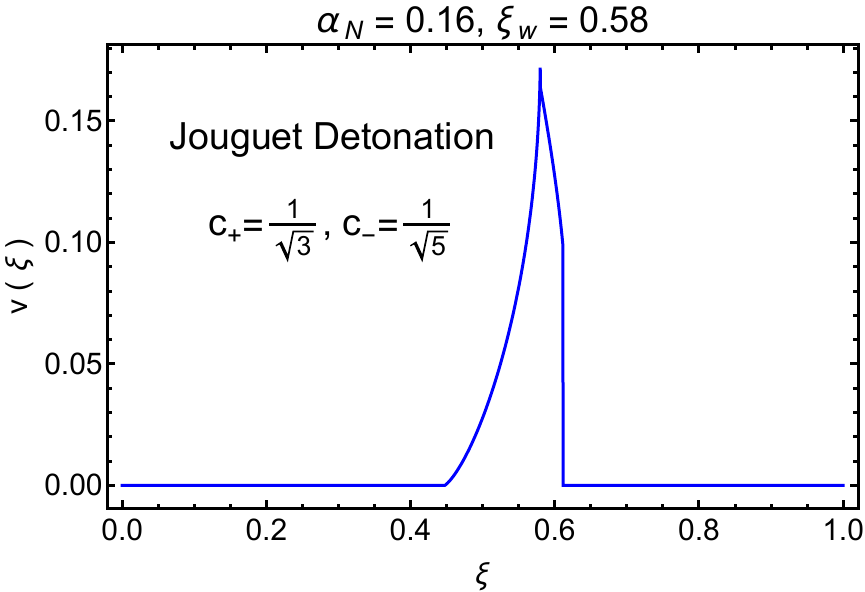}
\includegraphics[width=0.49\textwidth]{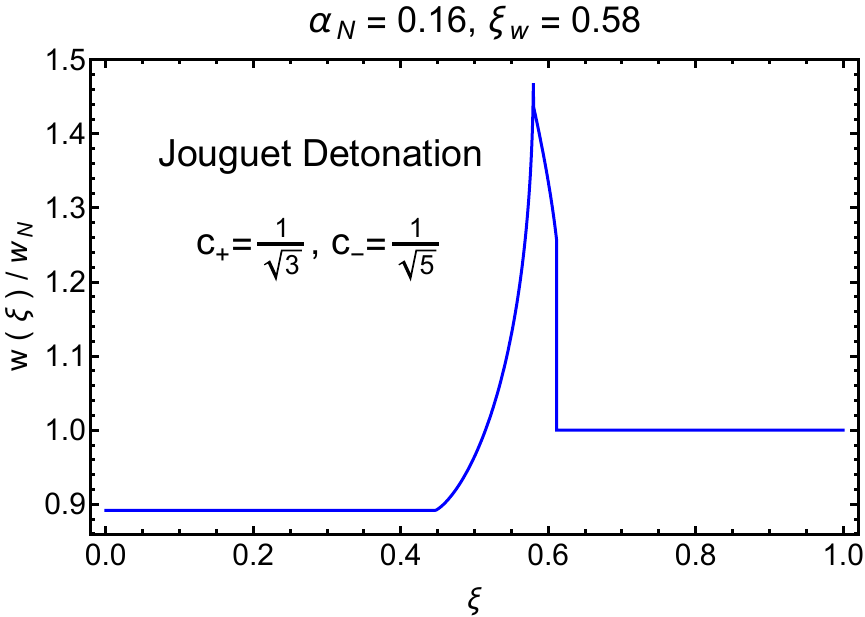}\\
\caption{The fluid profiles $v(\xi)$ (left) and $w(\xi)/w_N$ (right) of the Jouguet detonation mode. We fix $\alpha_N=0.16$ and take $c_+=1/\sqrt{3}$, $c_-=1/\sqrt{5}$. The blue lines correspond to the bubble wall velocity $\xi_w$ taking the value of 0.58.}
\label{fig:JouguetDetona}
\end{figure*}

The Jouguet detonation mode (absent in the bag model) is a special detonation mode ($\overline{v}_+>\overline{v}_-$) of Jouguet type ($\overline{v}_-=c_-$) corresponding to the plus-sign branch of~\eqref{eq:vbarpm} realized by $\overline{v}_+=v_{J}^{\rm det}(\alpha_+)$. Similar to the Jouguet deflagration mode, the fluid velocity profile of Jouguet detonation mode in Fig.~\ref{fig:JouguetDetona} also contains both compressive shockwave and rarefraction wave in the front and back of the bubble wall, respectively, but corresponding to the purple region in Fig.~\ref{fig:VelocityProfile}. To derive the fluid velocity profile, the Jouguet detonation condition $\overline{v}_{-}=c_-$ leads to $\overline{v}_+=v_{J}^{\rm det}$ from~\eqref{Jouguet:detonation} by~\eqref{eq:vbarpm}, giving rise to $v(\xi_w+0^{-})\equiv v_+=\mu(\xi_w,c_-)$ and $v(\xi_w+0^{+})\equiv v_-=\mu(\xi_w,v_{J}^{\rm det})$ that can be used to solve the fluid EoM~\eqref{eq:dv} both forward and backward from $(\xi_w+0^+, v_+)$ and $(\xi_w+0^-,v_-)$ with $c_s=c_+$ and $c_s=c_-$, respectively. The solved velocity profile again vanishes in front of the shockwave front $\xi=\xi_{sh}+0^+$ and behind $\xi=c_-+0^-$ as in the weak deflagration and weak detonation cases. The enthalpy profile can be similarly obtained from integrating~\eqref{eq:dw} backward from both $w(\xi_{sh}+0^-)=w_L$ and $w(\xi_w+0^-)=w_-$, where the enthalpies $w_L$ and $w_-$ are sequentially determined by the junction conditions~\eqref{eq:ShockJunction1} and~\eqref{eq:WallJunction1} with $w_R=w_N$, $\tilde{v}_R=\xi_{sh}$, $\tilde{v}_L=\mu(\xi_{sh},v_{sh})$ and $w_+=w(\xi_w+0^+)$, $\overline{v}_+=v_{J}^{\rm det}$, $\overline{v}_-=c_-$, respectively.

Finally, we discuss the condition where the Jouguet detonation mode can be realized. The difference between the weak detonation and Jouguet detonation mode is that the Jouguet detonation mode has a compressive shockwave in front of the wall. From the analysis of the weak deflagration mode, we can figure out that the situation where a shockfront can exist is $\mu(\xi_{sh},v_{sh})\xi_{sh}<c_+^2$, corresponding to the red and purple regions in Fig.~\ref{fig:VelocityProfile}. If the condition cannot be satisfied even at $\xi_{sh}=\xi_w$, the compressive shockwave must vanish and only weak detonation mode exists. Since just right in front of the wall $\mu(\xi_{w},v_+)$ takes the value of $v_{J}^{\rm det}$, we can derive that the form of the condition $\mu(\xi_{sh},v_{sh})\xi_{sh}<c_+^2$ turns into $v_{J}^{\rm det}\xi_w<c_+^2$ at $\xi_{sh}=\xi_w$, leading to $\xi_w<c_+^2/v_{J}^{\rm det}$. Recall that for a detonation mode $\overline{v}_-<\overline{v}_+<\xi_w$, we must have $\xi_{w}>\overline{v}_+|_{\rm min}=v_{J}^{\rm det}$. Hence, the existence of both detonation profile as wall as the shockwave can only be realized when $v_{J}^{\rm det}<\xi_w<c_+^2/v_{J}^{\rm det}$, that is exactly the condition where the Jouguet detonation mode can be realized. Note that if $c_+<v_{J}^{\rm det}$, the condition $v_{J}^{\rm det}<\xi_w<c_+^2/v_{J}^{\rm det}$ cannot be satisfied at all. Therefore, the condition of the realization of the Jouguet detonation can be concluded as
\begin{equation}
    \begin{split}
        v_{J}^{\rm det}<\xi_w<\frac{c_+^2}{v_{J}^{\rm det}},\ \ \ &\text{if}\ \  c_+>v_{J}^{\rm det}\\
        \text{no Jouguet detonation mode},\ \ \ &\text{if}\ \ c_{+}<v_{J}^{\rm det}.
    \end{split}
\end{equation}
Recall that we have shown the minimum of $v_{J}^{\rm det}(\alpha_+)$ is $c_-$, then when $c_+<c_-$, the condition $c_+>v_{J}^{\rm det}$ cannot be satisfied at all and the purple region will vanish in Fig.~\ref{fig:VelocityProfile}. Therefore, the Jouguet detonation mode can exist only in the $c_+>c_-$ case.

\bibliography{ref}

\end{document}